# A Compiler from Array Programs to Vectorized Homomorphic Encryption

ROLPH RECTO, Cornell University, USA
ANDREW C. MYERS, Cornell University, USA

Homomorphic encryption (HE) is a practical approach to secure computation over encrypted data. However, writing programs with efficient HE implementations remains the purview of experts. A difficult barrier for programmability is that efficiency requires operations to be *vectorized* in inobvious ways, forcing efficient HE programs to manipulate ciphertexts with complex data layouts and to interleave computations with data movement primitives.

We present Viaduct-HE, a compiler generates efficient vectorized HE programs. Viaduct-HE can generate both the operations and complex data layouts required for efficient HE programs. The source language of Viaduct-HE is *array-oriented*, enabling the compiler to have a simple representation of possible vectorization schedules. With such a representation, the compiler searches the space of possible vectorization schedules and finds those with efficient data layouts. After finding a vectorization schedule, Viaduct-HE further optimizes HE programs through term rewriting. The compiler has extension points to customize the exploration of vectorization schedules, to customize the cost model for HE programs, and to add back ends for new HE libraries.

Our evaluation of the prototype Viaduct-HE compiler shows that it produces efficient vectorized HE programs with sophisticated data layouts and optimizations comparable to those designed by experts.

## 1 Introduction

Homomorphic encryption (HE), which allows computations to be performed on encrypted data, has recently emerged as a viable way for securely offload computation. Efficient libraries [Chen et al. 2017] and hardware acceleration [Reagen et al. 2021; Riazi et al. 2020] have improved performance to be acceptable for practical use in a diverse range of applications such as the Password Monitor in the Microsoft Edge web browser [Lauter et al. 2021], privacy-preserving machine learning [Gilad-Bachrach et al. 2016], privacy-preserving genomics [Kim et al. 2021], and private information retrieval [Menon and Wu 2022].

Writing programs to be executed under HE, however, remains a forbidding challenge [Viand et al. 2021]. In particular, modern HE schemes support data encodings that allow for single-instruction, multiple data (SIMD) computation with very long vector widths but limited data movement capability.[1] SIMD parallelism allows developers to recoup the performance loss of executing programs in HE, but taking advantage of this capability requires significant expertise: efficient vectorized HE programs requires carefully laying out data in ciphertexts and interleaving data movement operations with computations. There is a large literature on efficient, expert-written vectorized HE implementations [Aharoni et al. 2023; Brutzkus et al. 2019; Gilad-Bachrach et al. 2016; Juvekar et al. 2018; Kim et al. 2023].

Prior work has developed compilers to ease the programmability burden of HE, but most work has targeted specific applications [Aharoni et al. 2023; Boemer et al. 2019a,b; Dathathri et al. 2019; Malik et al. 2021; van Elsloo et al. 2019], or focuses on challenges other than vectorization [Archer et al. 2019; Crockett et al. 2018; Dathathri et al. 2020; Lee et al. 2022]. Some HE compilers do attempt to generate vectorized implementations for arbitrary programs, but either fix simple data layouts for all programs [Viand et al. 2022] or require users to provide at least *some* information about complex data layouts [Cowan et al. 2021; Malik et al. 2023].

---

[1]Also known as ciphertext "packing" or "batching" in the literature.

We make the important observation that the complex, expert-written data layouts targeting specific applications in prior work are made possible by *array-level* reasoning. That is, given an array as input to an HE program, searching for an efficient layout amounts to asking such questions as "should this dimension of the array be vectorized in a single ciphertext, or be exploded along multiple ciphertexts?" This kind of reasoning is not reflected in the prior work on HE compilers, but as we show, it enables vectorized HE implementations with expert-level efficiency.

With this in mind, we reframe the problem of compiling efficient vectorized HE programs as two separate problems. First, a program must be "tensorized" and expressed as an *array program*. In many cases, this step is actually unnecessary, since the program can already be naturally expressed as operations over arrays. This is true for many HE applications, such as secure neural network inference. Once expressed as a computation over arrays, the space of possible vectorization schedules for the program can be given a simple, well-defined representation. This makes the "last-mile" vectorization of array programs much more tractable than the vectorization of arbitrary imperative programs.

This "tensorize-then-vectorize" approach is arguably already present in the literature. For example, Malik et al. [2021] developed an efficient vectorized HE implementation for evaluating decision forests by expressing the evaluation algorithm as a sequence of element-wise array operations and matrix–vector multiplication, and then using an existing kernel [Halevi and Shoup 2014] to implement matrix–vector multiplication efficiently.

In this paper, we aim to tackle the challenge of generating vectorized HE implementations for array programs. To this end, we propose Viaduct-HE, a vectorizing HE compiler for an array-oriented source language. Viaduct-HE simultaneously generates the complex data layouts and operations required for efficient HE implementations. Unlike in prior work [Cowan et al. 2021; Malik et al. 2023], this process is completely automatic: the compiler needs no user hints to generate complex layouts. The compiler leverages the high-level array structure of source programs to give a simple representation for possible vectorization schedules, allowing it to efficiently explore the space of schedules and find efficient data layouts. Once a schedule has been found, the compiler can further optimize the program by translating it to an intermediate representation amenable to term rewriting.

Viaduct-HE is designed to be extensible: after optimization, the compiler translates circuits into a loop nest representation designed for easy translation into operations exposed by HE libraries, allowing for the straightforward development of back ends that target new HE implementations. The compiler also has well-defined extension points for customizing the exploration of vectorization schedules and for estimating the cost of HE programs.

We make the following contributions:

- A high-level array-oriented source language that can express a wide variety of programs that target vectorized homomorphic encryption.
- New abstractions to represent vectorization schedules and to control the exploration of the schedule search space for array-oriented HE programs, facilitating automated search for efficient data layouts.
- Intermediate representations that enable optimization through term rewriting and easy development of back ends for new HE libraries.
- A prototype implementation of the Viaduct-HE compiler and an evaluation that shows that the prototype automatically generates efficient vectorized HE programs with sophisticated layouts comparable to those developed by experts.

## 2 Background on Homomorphic Encryption

Homomorphic encryption schemes allow for operations on ciphertexts, enabling computations to be securely offloaded to third parties without leaking information about the encrypted data. Such schemes are *homomorphic* in that ciphertext operations correspond to plaintext operations: given encryption and decryption functions **Enc** and **Dec**, for a function $f$ there exists a function $f'$ such that $f(x) = \textbf{Dec}(f'(\textbf{Enc}(x)))$.

In a typical setting involving homomorphic encryption, a client encrypts their data with a private key and sends the ciphertext to a third-party server. The server performs operations over the ciphertext, and then sends the resulting ciphertext back to the client. The client can then decrypt the ciphertext to get the actual result of the computation.

We target modern lattice-based homomorphic encryption schemes such as BFV [Fan and Vercauteren 2012], BGV [Brakerski et al. 2014], and CKKS [Cheon et al. 2017]. In these schemes, ciphertexts can encode many data elements at once. Thus we can treat ciphertexts as *vectors* of data elements. Homomorphic computations are expressed as addition and multiplication operations over ciphertexts. Addition and multiplication execute element-wise over encrypted data elements, allowing for SIMD processing: given ciphertexts $x = \textbf{Enc}([x_1, x_2, \ldots, x_n])$ and $y = \textbf{Enc}([y_1, y_2, \ldots, y_n])$, homomorphic addition $\oplus$ and multiplication $\otimes$ operate such that

$$\textbf{Dec}(x \oplus y) = [x_1 + y_1, x_2 + y_2, \ldots, x_n + y_n] \qquad \textbf{Dec}(x \otimes y) = [x_1 \times y_1, x_2 \times y_2, \ldots, x_n \times y_n].$$

There are analogous addition and multiplication operations between ciphertexts and plaintexts, which also have a vector structure. This allows computation over ciphertexts using data known to the server. For example, it is common to multiply a ciphertext with a plaintext *mask* consisting of 1s and 0s to zero out certain slots of the ciphertext.

Along with addition and multiplication, *rotation* facilitates data movement, cyclically shifting the slots of data elements by a specified amount. For example,

$$\textbf{Dec}(\text{rot}(-1, x)) = [x_2, x_3, \ldots, x_n, x_1] \qquad \textbf{Dec}(\text{rot}(2, x)) = [x_{n-1}, x_n, x_1, \ldots, x_{n-2}].$$

### 2.1 Programmability Challenges

While vectorized homomorphic encryption presents a viable approach to secure computation, there are many challenges to developing programs that use it. Such challenges include the lack of support for data-dependent control flow that forces programs to be written in "circuit" form; the selection of cryptographic parameters that are highly sensitive to the computations being executed; the management of ciphertext noise; and the interleaving of low-level "ciphertext maintenance" operations with computations [Viand et al. 2021].

Here we focus on the challenge of writing vectorized programs that use the SIMD capability of HE schemes. Efficiently vectorized HE programs are very different from programs in other regimes supporting SIMD. We now highlight some of the novelties of vectorizing in the HE regime.

***Very long vector widths.*** Vector widths in HE ciphertexts are large powers of two—on the order of thousands when the scheme's parameters are set to appropriate security levels [Albrecht et al. 2018]. To take advantage of such a large number of slots, often times HE programs are structured in counterintuitive ways. For example, the convolution kernel in Gazelle [Juvekar et al. 2018] applies a filter to all output pixels simultaneously. COPSE [Malik et al. 2021] evaluates decision forests by evaluating *all* branches at once and then applies masking to determine the right classification label for an input.

***Limited data movement.*** Although ciphertexts can be treated as vectors, they have a very limited interface. In particular, one cannot index into a ciphertext to retrieve individual data elements. All operations are SIMD and compute on entire ciphertexts at once. Thus expressing computation that operates on individual data elements as ciphertext operations can be challenging. One might consider naive approaches to avoid such difficulties; for example, ciphertexts can be treated as single data elements by only using their first slot. Failure to restructure programs to take advantage of the SIMD capability of HE, however, exacts a steep performance hit: in many cases, orders of magnitude in slowdown [Viand et al. 2022].

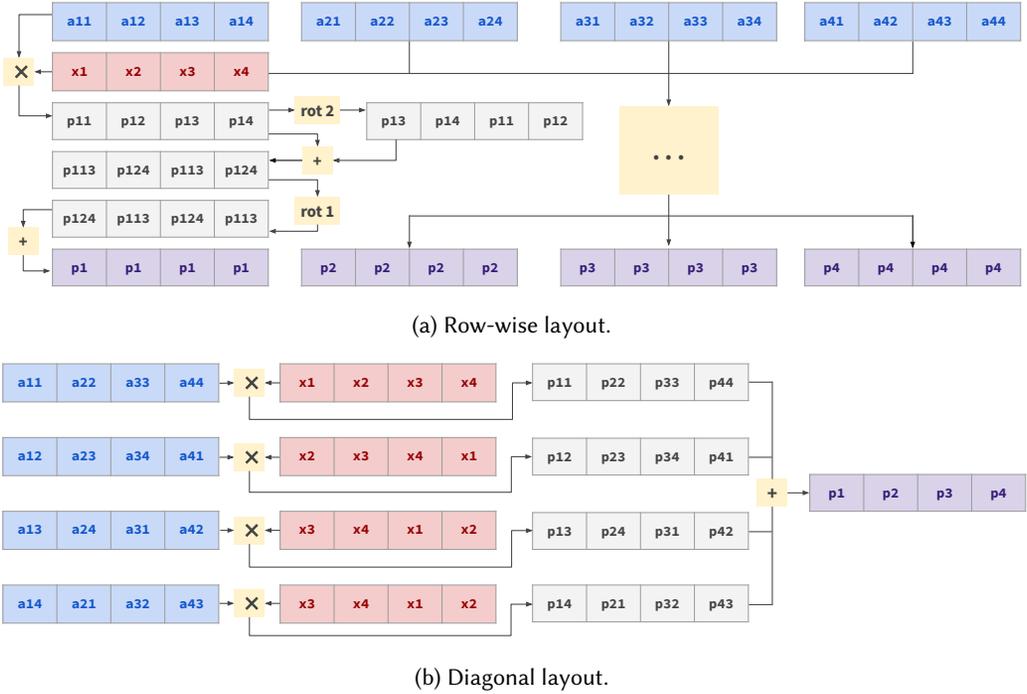

Fig. 1. Different layouts for matrix–vector multiplication. Vectors from matrix a are in blue; vectors from vector v are in red; the output vectors are in violet.

So in practice, data elements must be packed in ciphertexts to write efficient HE programs. However, packing creates new problems: if an operation requires data on different slots, ciphertexts must be rotated to align the operands. One is thus forced to interleave data movement and computation, but determining how to schedule these together efficiently can be difficult.

Because rotation operations provide limited data movement, the initial data layout in ciphertexts has a great impact on the efficiency of HE programs. One layout might aggressively pack data to minimize the number of ciphertexts the client needs to send and also minimize the computations the server needs to perform, but might require too many rotations; another layout might not aggressively pack data into ciphertexts to avoid the necessity of data movement operations, but might force greater client communication and the server to perform more computations.

***Example: Matrix–vector multiplication.*** To illustrate the challenges of developing vectorized HE programs, consider the two implementations of matrix–vector multiplication shown in Figure 1. Here a matrix a is multiplied with a vector x; the vectors containing data elements from a are in blue, while the vectors containing data elements from x are in red; the output vectors of the multiplication are in purple.

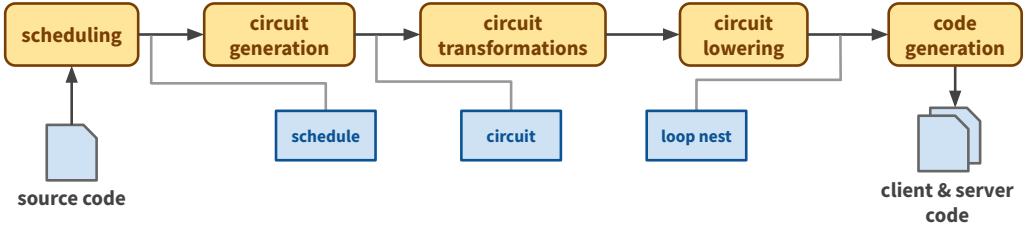

Fig. 2. Viaduct-HE compiler architecture.

Figure 1a shows a row-wise layout for the program, where the each of the blue vectors represents a row from a. A single red vector contains the vector x. The figure shows the computation of a dot product for one row of a and x; First, the vectors are multiplied, and then the product vector is rotated and added with itself multiple times to compute the sum. This pattern, which we call *rotate-and-reduce*, is common in the literature [Cowan et al. 2021; Kim et al. 2023; Viand et al. 2022] and it exploits it allows for computing reductions in a logarithmic number of operations relative to the number of elements.[2] Here 4 elements can be summed with 2 rotations and 2 additions. The row-wise layout results in the dot product outputs to be spread out in 4 ciphertexts, which can preclude further computation (they cannot be used as a packed vector in another matrix–vector multiplication, say) and induce a lot of communication if the server sends these outputs to the client.

Figure 1b shows the "generalized diagonal" layout from Halevi and Shoup [2014]. Here the vectors contain diagonals from array a: the first vector contains the main diagonal; the second vector contains a diagonal shifted to the right and wrapped around; and so on. These vectors are then multiplied with the vector containing x, but rotated an appropriate number of slots. The layout of the product vectors allow the sum to be computed simply by adding the vectors together, as the product elements for different rows are packed in the same vector but elements for the same row are "exploded" along multiple vectors.

In total, the diagonal layout requires only 3 rotations and 3 additions, compared to the 8 additions and 8 rotations required by the row-wise layout. Additionally, the outputs are packed in a single vector, which can be convenient for further computations (it can be used as input to another matrix–vector multiplication) or for returning results to the client with minimal communication.

## 3 Compiler Overview

Figure 2 shows the architecture of the Viaduct-HE compiler. It takes as input an array program and generates both code run by the client, which sends inputs and receives program results, and by the server, which performs the computations that implement the source program. The compiler has well-defined extension points to control different aspects of the compilation process. To describe this process in detail, we consider the compilation of a program that computes the distance of a client-provided point (x) against a list of test points known to the server (a).

### 3.1 Source Language

Figure 3a shows the source code for the distance program. It specifies that a is a 2D array provided as input by the server, with an extent of 4 on both dimensions; similarly, x is a 1D array provided as input by the client, with an extent of 4. Thus a is assumed to be known to the server and thus is in

---

[2]The pattern is sometimes called "rotate-and-sum," but it clearly also applies to products as well. It requires the dimension size to be a power of two and the reduction operator to be associative and commutative, which is true for addition and multiplication.

```
input a: [4, 4] from server
input x: [4] from client          at    = PlaintextArray([4])
for j: 4 {                        tmp0  = Vec(a,Roll(1,0),[0,0],
  sum(for i: 4 {                          [(2,0,0,{(0,1)}])
    (a[j][i] - x[i]) *            at[0] = encode(tmp0)
    (a[j][i] - x[i])              ...
  })                              xt    = CiphertextArray([4])
}                                 xt[0] = Vec(x,Id,[0],[(4,0,0,{(0,1)})])
                                  ...
         (a) Source code.         out   = CiphertextArray([])
let out =                         for i in range(4) {
  sum_vec(i: 4,                     inst1 = rot(C,i,xt[i])
    (at - rot(i, xt)) *             inst2 = sub(CP,inst1,at[i])
    (at - rot(i, xt)))              inst3 = mul(C,inst2,inst2)
                                    out   = add(C,out,inst3)
    (b) Circuit representation.   }
```

(c) Loop-nest representation.

Fig. 3. Compilation of the distance program.

plaintext, while x is in ciphertext since it comes from the client. The two **for** nodes each introduce a new dimension to the output array; they also introduce the index variables i and j, which are used to index into the input arrays a and x. The dimension introduced by the inner **for** node is reduced with the **sum** operator, so the output array has one dimension. Conceptually, the program computes the distance of x from the rows of a, each of which represents a point. An equivalent implementation in a traditional imperative language would look like the following program on the left.

```
a = Array[4][4];
x = Array[4];
out = Array[4]
for (j = 0; j < 4; j++) {
  acc = 0;
  for (i = 0 ; i < 4; i++) {
    acc += (a[j][i] - x[i]) *
           (a[j][i] - x[i]);
  }
  out[j] = acc;
}
```

Figure 4 defines the abstract syntax for the source language. Programs consist of a sequence of inputs and let-bound expressions followed by an output expression whose result the server sends to the client. Expressions uniformly denote arrays; scalars are considered zero-dimensional arrays. The expression **input** $a : sh$ **from** $pt$ **in** $e$ denotes an array with shape $sh$ received as input from $pt$, which is either the client or the server. Input arrays from the client are treated as ciphertexts, while input arrays from the server are treated as plaintexts. Operation expression $e_1 \odot e_2$ denotes an element-wise operation over equal-dimension arrays denoted by $e_1$ and $e_2$, while reduction expression **reduce**$_{\odot,n}(e)$ reduces the $n$-th dimension of the array denoted by $e$ using $\odot$.

The expression **for** $i : n \{e\}$ adds a new outermost dimension with extent $n$ to the array denoted by $e$, while $e[in]$—also referred throughout as an *indexing site*—indexes the outermost dimension of an array. Only array variables, introduced by inputs or let-bindings, can be indexed. The compiler also imposes some restrictions on indexing expressions. Particularly, index variables cannot be

$$
\begin{array}{lll}
\text{Integer } z \in \mathbb{Z} & \text{Natural } n \in \mathbb{N} & \text{Index variable } i, j, k \quad \text{Array variable } a, b, c \\
\text{Shape} & sh \;::=\; [n_1, \ldots, n_d] \\
\text{Party} & pt \;::=\; \textbf{client} \mid \textbf{server} \\
\text{Operator} & \odot \;::=\; + \mid - \mid \times \\
\text{Index} & in \;::=\; i \mid z \mid in \odot in \\
\text{Expression} & e \;::=\; z \mid ie \mid e \odot e \mid \textbf{reduce}_{\odot, n}(e) \mid \textbf{for } i : n \;\{e\} \\
\text{Index Expression} & ie \;::=\; a \mid ie[in] \\
\text{Statement} & s \;::=\; \textbf{let } a = e \textbf{ in } s \mid \textbf{input } a : sh \textbf{ from } pt \textbf{ in } s \mid e \\
\end{array}
$$

Fig. 4. Abstract syntax for the source language.

multiplied together (e.g. a[i*j]), as compiler analyses assume that the dimensions of indexed arrays are traversed with constant stride.

## 3.2 Scheduling

The source program is an abstract representation of computation over arrays; it represents the *algorithm*—the *what*—of an HE program. The vectorization *schedule*—*how* data will be represented by ciphertext and plaintext and how computations will be performed by HE operations—is left unspecified by the source program. Because its source language is array-oriented, the vectorization schedules for Viaduct-HE programs have a simple representation, allowing the compiler to manipulate such schedules and search for efficient ones during its scheduling stage. The compiler provides extension points to control both how the search space of schedules is explored and how the cost of schedules are assessed.

Like matrix multiplication, the distance program can be given a row-wise layout and a diagonal layout. The diagonal layout similarly requires less rotation and addition operations. The scheduling stage of the compiler can search for these schedules and assess their costs.

## 3.3 Circuit Representation

Once an efficient schedule has been found, the circuit generation stage of the compiler uses it to translate the source program into a circuit representation. The circuit representation represents information about the ciphertexts and plaintexts required in an HE program, as well as operations to be performed over these, at a very abstract level. The compiler has circuit transformation stages that leverage the algeraic properties of circuits to rewrite them into more efficient forms. Circuits are designed to facilitate optimization: a single circuit expression can represent many computations, so circuit rewrites can optimize many computations simultaneously.

Figure 3b shows the circuit representation for the distance program with the diagonal layout. The `sum_vec` operation represents a summation of 4 different vectors together into one vector. The 4 vectors each represent a the result of a squared difference computation between vector containing a generalized diagonal of array a (represented by the variable at) and a rotated vector containing array v (represented by the variable xt).

## 3.4 Loop-nest Representation

Circuits represent HE computations at a very high level. This makes the circuit representation amenable to optimization, but makes generation of target code difficult. After circuit programs have been optimized, the circuit lowering stage of the compiler translates circuit programs into a "loop-nest" representation. Loop-nest programs are imperative programs that are much closer in structure to target code. Once in the loop-nest representation, the code generation stage of the compiler generates target code using a *back end* for a specific HE library. The compiler can generate code for a different HE library just by swapping out the back end it uses. Back ends only need to translate loop-nest programs to target code, so adding support for new back ends is straightforward.

Figure 3c shows the loop-nest representation for the distance program. It contains code to explicitly fill in the variables at and xt with the vectors that will be used in computations. The summation is now represented as an explicit **for** loop that accumulates squared distance computations in an out variable.

## 4 Scheduling

The scheduling stage begins by first translating source programs into an intermediate representation that eliminates explicit indexing constructs. From there, the compiler generates an initial schedule and explores the search space of vectorization schedules.

### 4.1 Index-free Representation

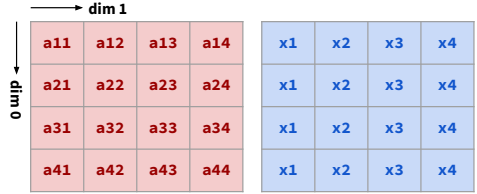

The index-free representation is similar to the source language, except that **for** nodes are eliminated and indexing sites are replaced with pair (*is*, *at*) of a unique identifier (*is*) and an *array traversal* (*at*) that summarizes the contents of the array denoted by the indexing site. Array traversals are arrays generated from indexing another array. The figure on the left shows the array traversals in the distance program. The traversal in red is from indexing array a; the traversal in blue is from indexing array x. Note that the dimension 0 is introduced by the **for** j node in the source program, while dimension 1 is introduced by the **for** i node. The traversal of array x repeats along dimension 0 because it is not indexed by j and thus does not change along that dimension.

Formally, array traversals have three components: the name of the indexed array; the integer *offsets* at which the traversal begins, defined by a list of integers with a length equal to the number of dimensions of the indexed array; and a list of *traversal dimensions* (*td*). We write $a(z_1, \ldots, z_m)[td_1, \ldots, td_n]$ to denote a $n$-dimensional traversal of an $m$-dimensional array. Array traversals can define positions that are out-of-bounds; for example, offsets can be negative even though all index positions in an array start at 0.

Each traversal dimension has an *extent* specifying its size and a set of *content dimensions* that specify how the dimensions traverses the indexed array. Content dimensions have a *dimension index* and a *stride*. For example, a traversal dimension (4, {0 :: 2}) defines a traversal of an array along its zeroth dimension that spans 4 elements, where only every other element is traversed (i.e. the stride is 2). Traversal dimensions can have empty content dimension sets, which means that the array traversal does not vary along the dimension. We call these traversal dimensions *empty*.

For example, the index-free representation for the distance program is

**sum**(1, ((at1,at) - (xt1,xt)) * (((at2,at) - (xt2,xt))).

Variables at1 and at2 represent indexing sites with traversal at that indexes array a; variables xt1 and xt2 represent indexing sites with traversal xt that indexes array x. The array traversals denoted by these indexing sites is as follows:

$$at = a(0,0)[(4, \{0 :: 1\}), (4, \{1 :: 1\})] \qquad xt = x(0)[(4, \{\}), (4, \{0 :: 1\})].$$

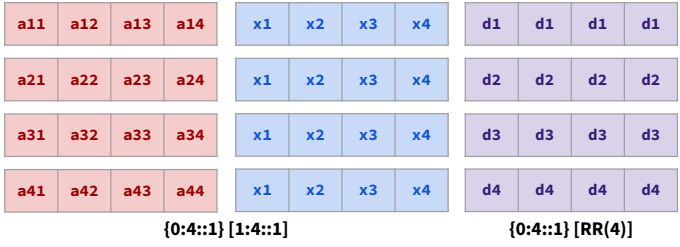

(a) Row-wise layout. Induces 5 input vectors, 4 output vectors, 8 additions, 8 rotations (2 adds and rotates per vector with rotate-and-reduce).

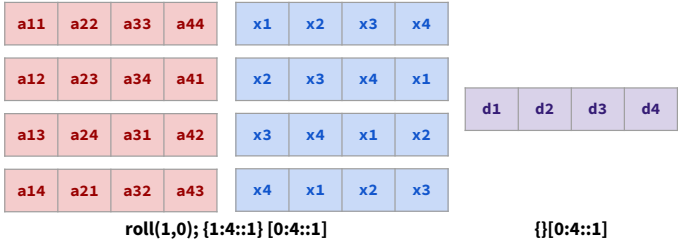

(b) "Diagonal" layout. Induces 5 input vectors, 1 output vector, 3 additions, 3 rotations.

Fig. 5. Different schedules for the distance program.

The traversal at defines a 4x4 array where dimension 0 traverses dimension 0 of input array a with stride 1, and dimension 1 traverses dimension 1 of array a with stride 1. Meanwhile, the traversal xt also defines a 4x4 array, but its dimension 0 is empty and its dimension 1 traverses the only dimension of input array x with stride 1.

### 4.2 Representing Schedules

Schedules define a *layout* for the array traversals denoted by each indexing site in the index-free representation. The layout determines how an array traversal is represented as a set of vectors. One can think of layouts as a kind of traversal of array traversals themselves. Because of this, layouts are defined similarly to array traversals, except they do not specify offsets, as layouts always have offset 0 along every traversal dimension.

Layouts are built from *schedule dimensions* which denote some part of an array traversal. We write the syntax $i : n :: s$ for a schedule dimension with dimension index $i$, extent $n$, and stride $s$. For example, a schedule dimension $0 : 4 :: 2$ defines a 4-element section of an array traversal along its zeroth dimension that contains only every other element (i.e. the stride is 2).

Concretely, layouts consist of the following: (1) a set of *exploded* dimensions; (2) a list of *vectorized* dimensions; and (3) a preprocessing operation. Exploded dimensions define parts of the array traversal that will be laid out in *different* vectors, while vectorized dimensions define parts that will be laid out in *every* vector. The ordering of vectorized dimensions defines their ordering on a vector: the beginning of the list defines the outermost vectorized dimensions, while the end defines the innermost vectorized dimensions. Preprocessing operations change the contents of the array traversal before being laid out into vectors, which allow for the representation of complex layouts. We write $p\{ed_1, \ldots, ed_m\}[vd_1, \ldots, vd_n]$ to denote a schedule with preprocessing $p$, $m$ exploded dimensions ($ed$) and $n$ vectorized dimensions ($vd$). When $p$ is the identity preprocessing operation, it is often omitted from the schedule.

Figure 5 shows two different schedules for the distance program. The vectors of at1 and at2 are in red, while the vectors of xt1 and xt2 are in blue. Their respective layouts are given below the vectors. Finally, the vectors of the distance program's output is in purple, and the output layout is given below. Note that array traversals for at and xt must have the same layout since their arrays are multiplied together, and operands of element-wise operations must have the same layout.

Figure 5a represents a row-wise layout, where the entirety of dimension 1 of at and xt are vectorized while the entirety dimension 0 is exploded into multiple vectors. Thus traversal at is represented by 4 vectors, one for each of its rows; since traversal xt has 4 equal rows, it is represented by a single vector. Meanwhile, Figure 5b represents a "diagonal" layout; it is similar to a column-wise layout where each vector contains a column, but the roll preprocessing operation rotates the rows along the columns, where the rotation amount progressively increases. As discussed in §2, a similar diagonal layout was originally specified in Halevi and Shoup [2014] as an efficient implementation of matrix–vector multiplication, but here we see that the schedule abstraction can capture its essence, allowing the compiler to generalize and use it for other programs.

Note that exploded dimensions have a name associated with them; in the above syntax, the name for exploded dim $i$ is $d_i$. These names are used to uniquely identify vectors induced by the layout. During circuit generation, the names of exploded schedule dimensions will be used as variables that parameterize circuit expressions.

***Preprocessing.*** A preprocessing operation in a layout transforms an array traversal before laying it out into vectors. Formally, a preprocessing operation is a permutation over elements of the array traversal. Thus we can think of preprocessing operations as functions from element positions to element positions. Given an $n$-dimensional array traversal $at$, applying preprocessing operation $p$ over $at$ defines a new traversal such that the element at position $x_1, \ldots, x_n$ is the element at position $p(x_1, \ldots, x_n)$ of $at$. For example, the identity preprocessing operation is the trivial permutation that maps element positions to themselves: $\text{id} = \lambda(x_1, \ldots, x_n).(x_1, \ldots, x_n)$. Given that both dimensions $i$ and $j$ of an array traversal both have extent $n$, we can define the roll preprocessing operation as follows:

$$\text{roll}(a, b) = \lambda(x_1, \ldots, x_a, \ldots, x_b, \ldots, x_n).(x_1, \ldots, x_a + x_b \mathbin{\%} n, \ldots, x_b, \ldots x_n).$$

***Applying layouts.*** When applied to an array traversal, a layout generates a set of vectors that contain parts of the array indexed by the traversal. Formally, a vector contains four components: the name of the indexed array; a preprocessing operation; a list of integer offsets; and a list of traversal dimensions. As with preprocessing in a layout, a preprocessing operation in a vector transforms an array before its contents are laid out in the vector. We write $a.p(z_1, \ldots, z_m)[vtd_1, \ldots, vtd_n]$ to denote a vector indexing an $m$-dimensional array $a$ with preprocessing $p$ and $n$ traversal dimensions ($vtd_i$). Again $p$ is usually elided when it is the identity preprocessing operation. Vector traversal dimensions are similar to array traversal dimensions, except they also track elements that are out-of-bounds. A vector can have out-of-bounds values either because its dimensions extend beyond the extents of the indexed array or the extents of the array traversal from which it is generated. We write $(n, obl, obr, \{cd_1, \ldots, cd_m\})$ to denote a vector traversal dimension with extent $n$, $m$ content dimensions, a left out-of-bounds extent $obl$, and a right out-of-bounds extent $obr$. The left out-of-bounds and right out-of-bounds extents count the number of positions in a dimension that are out-of-bounds to the left and right of the in-bounds positions respectively. The compiler enforces the semantics that out-of-bounds values have value 0.

Given a layout with $n$ exploded dimensions each with extent $n_i$, applying the layout to an array traversal generates $\prod n_i$ vectors, one for each distinct combination of positions that can be defined along exploded dimensions. We call each such combination a *coordinate*. For example, when applied to array traversal at, the diagonal layout for the distance program generates 4 vectors, one for each distinct positions that the exploded dimension named $i$ can take:

$\{i \mapsto 0\} \mapsto \text{a.roll}(1,0)(0,0)[(4,0,0,\{0::1\})]$    $\{i \mapsto 1\} \mapsto \text{a.roll}(1,0)(0,1)[(4,0,0,\{0::1\})]$

$\{i \mapsto 2\} \mapsto \text{a.roll}(1,0)(0,2)[(4,0,0,\{0::1\})]$    $\{i \mapsto 3\} \mapsto \text{a.roll}(1,0)(0,3)[(4,0,0,\{0::1\})]$.

This represents the same vectors for traversal at visually represented in Figure 5b.

## 4.3 Searching for Schedules

To search for an efficient schedule for the program, the scheduling stage begins with an initial schedule where the layouts for all indexing sites contain only exploded dimensions. Thus in this schedule elements of arrays are placed in individual vectors. While very inefficient, the initial schedule can be defined for any program. To explore the search space of schedules, the scheduling stage uses a set of *schedule transformers* that take a schedule as input and returns a set of "nearby" schedules. To assess both the validity of a schedule visited during search, the compiler attempts to generate a circuit from the schedule. If a circuit is successfully generated, it is applied to a *cost estimator* function to determine the cost of the schedule.

The schedule transformers in the prototype implementation of the compiler include the following.

***Vectorize dimension transformer.*** This transformer takes an exploded dimension from a layout and vectorizes it:

$$p\{\ldots, (d_a)\ i_a : n_a :: s_a, \ldots\}[\ldots] \rightsquigarrow p\{\ldots\}[\ldots, i_a : n_a :: s_a].$$

Importantly, this transformer only generates vectorized dimensions with extents that are powers of two; if the exploded dimension is not a power of two, the transformer will round up the vectorized dimension's extent to the nearest one. The transformer imposes this limit on vectorized dimensions to simplify reasoning about correctness: vectors only wrap around correctly when their size divides the slot counts of ciphertexts and plaintexts without remainder, and these slot counts are always powers of two. This limitation also allows the circuit generation stage to uniformly use the rotate-and-reduce pattern.

***Tiling transformer.*** This transformer takes an exploded dimension and *tiles* it into an outer dimension and an inner dimension. That is, given that extent $n_a$ can be split into $n$ tiles each of size $t$ (i.e., $e_a = tn$), it performs the following transformation:

$$p\{\ldots, (d_a)\ i_a : n_a :: s_a, \ldots\}[\ldots] \rightsquigarrow p\{\ldots, (d'_a)\ i_a : t :: s_a, (d''_a)\ i_a : n :: s_a t, \ldots\}[\ldots]$$

where $d'_i$ and $d''_i$ are fresh exploded dimension names.

***Roll transformer.*** This transformer applies a roll preprocessing operation to a layout:

$$\text{id}\{\ldots, (d_a)\ i_a : n :: s_a, \ldots\}[i_b : n :: s_b, \ldots] \rightsquigarrow \text{roll}(a,b)\{\ldots, (d_a)\ i_a : n :: s_a, \ldots\}[i_b : n :: s_b, \ldots].$$

The transformer only applies when dimension $a$ is exploded, dimension $b$ is the outermost vectorized dimension and their extents match. Other conditions must hold to use a layout with roll preprocessing; we defer to the supplementary materials for more details.

***Epochs.*** The search space for vectorization schedules is large. To control the amount of time that scheduling takes, the search is staggered into *epochs*. During an epoch, the configuration of schedule transformers is fixed such that only a subset of the search space is explored. When no more schedules can be visited, the epoch ends; a new epoch then begins with the schedule transformers updated to allow exploration of a bigger subset of the search space. The compiler runs a set number of epochs, after which it uses the most efficient schedule found to proceed to later stages of compilation.

| | | | |
|---|---|---|---|
| | Dimension variable $d \in \mathcal{D}$ | Array name $a$ | Vector $v$ |
| | Plaintext var $\varphi_p$ | Ciphertext var $\varphi_c$ | Offset var $\varphi_o$ |
| | Circuit value map | $cm_\tau \in (\mathcal{D} \times \cdots \times \mathcal{D} \rightharpoonup \mathbb{N} \times \cdots \times \mathbb{N}) \to \tau$ | |
| Offset | $oe$ | ::= | $d \mid z \mid oe \odot oe \mid \varphi_o$ |
| Expression | $ce$ | ::= | $\varphi_p \mid \varphi_c \mid z \mid ce \odot ce \mid \textbf{rot}(oe, ce) \mid \textbf{reduce-vec}_\odot(d : n, ce)$ |
| Statement | $cs$ | ::= | $\textbf{let } a : [d_1 : n_1, \ldots, d_n : n_n] = ce \textbf{ in } cs \mid \textbf{skip}$ |
| Object | $co \in O$ | ::= | $\text{Const}(z) \mid \text{Mask}((n, n, n)) \mid \text{Vec}(v)$ |
| Registry | $cr$ | ::= | $\varphi_o \mapsto cm_\mathbb{Z}, cr \mid \varphi_p \mapsto cm_O, cr \mid \varphi_c \mapsto cm_O, cr \mid \cdot$ |

Fig. 6. Abstract syntax for circuit programs.

The prototype implementation of the Viaduct-HE compiler uses epochs to control how schedule dimensions are split by the tiling transformer, which is the main cause of search space explosion. The tiling transformer gradually increases the number of schedule dimensions it splits as the number of scheduling epochs increase.

## 5 Circuit Generation

The circuit generation stage takes a schedule and index-free program as input and attempts to generate a circuit program. The design of the circuit representation reflects the fact that many computations in HE programs are structurally similar. Thus a circuit expression denotes not just a single HE computations, but rather a family of HE computations. Expressions are parameterized by *dimension variables*, and an expression represents a different computation for each combination of values (coordinates) these variables take.

Figure 6 shows the abstract syntax for circuit programs. A circuit program consists of a sequence of let statements that bind the results of expressions to array names; the last of these statements defines a distinguished output array (out) whose results will be sent to the client. A statement **let** $a : [d_1 : n_1, \ldots, d_n : n_n] = ce$ declares an array $a$ whose contents is computed by expression $ce$. Note that $ce$ is parameterized by dimension variables $d_i$ each with extent $n_i$, which means that $ce$ represents $\prod n_i$ different computations, one for each distinct combination of values that the variables can take. Because expressions can vary depending on the coordinates their dimension variables take, circuit programs are accompanied by a *circuit registry* data structure that records information about the exact values expressions take at a particular coordinate.

Circuit expressions include literals ($z$) and operations ($ce_1 \odot ce_2$) as in source programs. Expression **rot**($oe, ce$) rotates the vector denoted by $ce$ by an offset $oe$. Offsets can include literals, operations, index variables, and offset variables ($\varphi_o$); the latter two allows rotation amounts to vary depending on the values of in-scope dimension variables. The value of an offset variable at a particular coordinate is defined by a map that is stored in the registry that comes with the circuit program. Ciphertext ($\varphi_c$) and plaintext ($\varphi_p$) variables define a family of ciphertext and plaintext vectors respectively. Like offset variables, the exact vector these variables represent at a particular coordinates is defined by a map in the registry. These vectors can contain parts of input arrays and result arrays of prior expressions; additionally, plaintext variables can also represent constant vectors, which contain the same value in all of its slots, and mask vectors, which can be multiplied to another vector to zero out some of its slots. Masks are defined by a list of dimensions $[(n_1, lo_1, hi_1), \ldots, (n_n, lo_n, hi_n)]$, where $n_i$ is the extent of the dimension $i$ and $[lo_i, hi_i]$ is the *defined interval* for dimension $i$. The mask has value 1 in slots within defined intervals and 0 in slots outside of defined intervals.

Array Materializer $A$    Layout $\ell$    Index-free expression $fe$    Index-free statement $fs$
Schedule dimension $sd$    Exploded dimension $ed$
Output Vectorized Dimension    $ovd$    $::=$    $i:n::s \mid \text{R}(n) \mid \text{RR}(n)$
Output Layouts    $o\ell$    $::=$    $* \mid p\{ed_1, \ldots, ed_m\}[ovd_1, \ldots, ovd_n]$
Schedule    $\Sigma$    $::=$    $\Sigma, is : \ell \mid \cdot$
Input Context    $\Gamma$    $::=$    $\Gamma, a : sh \mid \cdot$
Expression Context    $\Delta$    $::=$    $\Delta, a : (sh, o\ell) \mid \cdot$

$$\boxed{A; \Sigma; \Gamma; \Delta \vdash fe \leadsto ce : (sh, o\ell)} \qquad \boxed{A; \Sigma; \Gamma; \Delta \vdash fs \leadsto cs}$$

$* <: sh \qquad * <: o\ell \qquad p\{\ldots\}[\text{RR}(n), ovd_2, \ldots, ovd_n] <: p\{\ldots\}[ovd_2, \ldots, ovd_n]$

CGen-Literal
$$\dfrac{}{A; \Sigma; \Gamma; \Delta \vdash z \leadsto z : (*, *)}$$

CGen-Op
$$\dfrac{A; \Sigma; \Gamma; \Delta \vdash fe_1 \leadsto ce_1 : (sh, o\ell) \qquad A; \Sigma; \Gamma; \Delta \vdash fe_2 \leadsto ce_2 : (sh, o\ell)}{A; \Sigma; \Gamma; \Delta \vdash fe_1 \odot fe_2 \leadsto ce_1 \odot ce_2 : (sh, o\ell)}$$

CGen-Input-Index
$$\dfrac{\Sigma(is) = \ell \qquad \Gamma(a) = sh_a \\ a = \text{tr-array}(at) \qquad sh = \text{tr-shape}(at) \\ \text{materialize-input}(A, sh_a, at, \ell) \leadsto ce}{A; \Sigma; \Gamma; \Delta \vdash (is, at) \leadsto ce : (sh, \ell)}$$

CGen-Expr-Index
$$\dfrac{\Sigma(is) = \ell \qquad \Delta(a) = (sh_a, o\ell_a) \\ a = \text{tr-array}(at) \qquad sh = \text{tr-shape}(at) \\ \text{materialize-expr}(A, sh_a, o\ell_a, at, \ell) \leadsto ce}{A; \Sigma; \Gamma; \Delta \vdash (is, at) \leadsto ce : (sh, \ell)}$$

CGen-Reduce
$$\dfrac{sh_1 = [n_0, \ldots, n_{n-1}, n_n, n_{n+1}, \ldots, n_{d-1}] \qquad sh_2 = [n_0, \ldots, n_{n-1}, n_{n+1}, \ldots, n_{d-1}] \\ A; \Sigma; \Gamma; \Delta \vdash fe \leadsto ce : (sh_1, o\ell_1) \qquad \text{reduce-layout}(n, o\ell_1) \leadsto (o\ell_2, dl)}{A; \Sigma; \Gamma; \Delta \vdash \mathbf{reduce}_{\odot, n}(fe) \leadsto \text{gen-reduce}_{\odot}(ce, dl) : (sh_2, o\ell_2)}$$

CGen-Input
$$\dfrac{A; \Sigma; \Gamma, a : sh; \Delta \vdash fs \leadsto cs}{A; \Sigma; \Gamma; \Delta \vdash \mathbf{input}\ a : sh\ \mathbf{from}\ pt\ \mathbf{in}\ fe \leadsto cs}$$

CGen-Let
$$\dfrac{o\ell = p\{(d_1)\ i_1 : n_1 :: s_1, \ldots, (d_n)\ i_n : n_n :: s_n\}[\ldots] \\ A; \Sigma; \Gamma; \Delta \vdash fe \leadsto ce : (sh, o\ell) \qquad A; \Sigma; \Gamma; \Delta, a : (sh, o\ell) \vdash fs \leadsto cs}{A; \Sigma; \Gamma; \Delta \vdash \mathbf{let}\ a = fe\ \mathbf{in}\ fs \leadsto \mathbf{let}\ a : [d_1 : n_1, \ldots, d_n : n_n] = ce\ \mathbf{in}\ cs}$$

Fig. 7. Rules for circuit generation.

Finally, the expression $\mathbf{reduce\text{-}vec}_{\odot}(d : n, ce)$ defines a computation where multiple vectors are reduced to a single vector with operation $\odot$. If the reduction expression is parameterized by dimension variables $d_1, \ldots, d_n$ with extents $n_1, \ldots, n_n$, then the expression represents $\prod n_i$ different vectors, each of which were computed by reducing $d$ vectors together. Thus the expression $ce$ is parameterized by variables $d_1, \ldots, d_n, d$.

## 5.1 Translation Rules for Circuit Generation

The translation into the circuit representation is mostly standard across programs, with the exception of the translation of indexing sites. The compiler uses a set of *array materializers* that lower the array traversals denoted by indexing sites into vectors and circuit operations according to a specific layout. We discuss them in detail in §5.2.

Figure 7 shows the rules for generating a circuit program from the index-free representation. The judgment defines both the translation to a circuit as well as the conditions that must hold for the translation to be successful. The expression translation judgment has form $A; \Sigma; \Gamma; \Delta \vdash fe \leadsto ce : (sh, \ell)$, which means that given an array materializer configuration $A$, schedule $\Sigma$, input context $\Gamma$, and expression context $\Delta$, the index-free expression $fe$ can be translated to circuit expression $ce$, where the computation defined by $ce$ has shape $sh$ and output layout $\ell$. The output layout defines how the results are laid out in vectors. The input context defines the shapes of input arrays in scope, while the expression context defines the shape and output layout of let-bound arrays in scope. The statement translation judgment $A; \Sigma; \Gamma; \Delta \vdash fs \leadsto cs$ has a similar form to expression translations. Additionally, judgment $sh_1 <: sh_2$ means that shape $sh_1$ can be coerced to shape $sh_2$, and similarly $o\ell_1 <: o\ell_2$ means that output layout $o\ell_1$ can be coerced to $o\ell_2$.

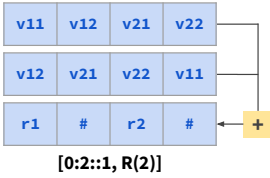

***Output layouts.*** Note that output layouts are more general than the layouts defined by schedules. First, they can be the "wildcard" layout ($*$), which can be coerced into any layout. Second, vectorized dimensions can take other forms. A *reduced dimension* ($R(n)$) represents a dimension in a vector with extent $n$, but since it is reduced only the first position of the dimension has an array element; the rest of the dimension contain invalid values. A vectorized dimension becomes a reduced dimension when its contents are rotated-and-reduced. The figure on the left shows the output layout for vector with a starting layout of $[0 : 2 :: 1, 1 : 2 :: 1]$ after its inner dimension has been reduced.

When the outermost vectorized dimension is rotated-and-reduced, however, the elements of the dimension wrap around such that the result of the reduction repeats along the extent of the dimension. This can be seen in the row-wise layouts for matrix–vector multiplication (Figure 1a) and the distance program (Figure 5a). In that case, a vectorized dimension with extent $n$ becomes a *reduced repeated* dimension ($RR(n)$). Output layouts with reduced repeated dimensions can be coerced into layouts that drop such dimensions.

***Translations.*** The translations of literals (CGen-Literal) and operations (CGen-Op) are straightforward; CGen-Op additionally ensures that the operands have the same shape and output layout. The translations for indexing sites (CGen-Input-Index and CGen-Expr-Index) use the compiler's array materializer configuration $A$ to lower an array traversal into a layout specified by the schedule. The functions tr-array and tr-shape return the indexed array and shape of an array traversal respectively; the functions materialize-input and materialize-expr are part of the interface of array materializers and, if successful, return a circuit expression representing the vectors of the array traversal in the required layout. The translations for statements (CGen-Input, CGen-Let, CGen-Output) add array information to the context. Note that the translation for let statements additionally uses the exploded dimensions of the output layout of its body expression circuit as dimension variables to parameterize the circuit.

The translation of reduction expressions (CGen-Reduce) are more involved. Given $\textbf{reduce}_{\odot,n}(fe)$ and that $fe$ is translated to $ce$, the output layout $o\ell_1$ of $ce$ is transformed to an output layout $o\ell_2$ that reflects the reduction by the reduce-layout function, which returns $o\ell_2$ layout as well as the list of schedule dimensions ($dl$) in $\ell_1$ that were reduced. Let $i$ be the traversal dimension index referenced by a schedule dimension $sd$ in $\ell_1$. Then there are three possible cases:
- When $i < n$, then $sd$ remains in $\ell_2$ unchanged.
- When $i > n$, then $sd$ remains in $\ell_2$ but now references traversal dimension index $i - 1$.
- When $i = n$, $sd$ is added to the list of reduced schedule dimensions $dl$. If $sd$ is exploded, it is removed from $\ell_2$ entirely; if $sd$ is vectorized with extent $n$, it is either replaced with a reduced dimension R($n$) or a reduced repeated dimension R($n$) depending on its position.

Note that REDUCE-LAYOUT fails when the preprocessing operation of the layout cannot be successfully transformed by the REDUCE-PREPROCESS function, which is specific to each preprocessing operation. Given identity preprocessing operation (id), REDUCE-PREPROCESS always succeeds and returns id unchanged. Meanwhile, given preprocessing roll($a, b$) and reduced dimension index $n$ REDUCE-PREPROCESS returns either id when $n = a$ or roll($a, b$) when $b \neq n \neq a$. When $n = b$, REDUCE-PREPROCESS is not defined and fails. Intuitively, reducing dimension $a$ transforms roll into id since it only changes the positions of elements along $a$. Meanwhile, reducing dimension $b$ would reduce array elements together that originally had positions with different values for $a$ before roll was applied, which is invalid.

Finally, the GEN-REDUCE$_\odot$ function generates the circuit expressions necessary to translate the reduction. It takes the list of reduced schedule dimensions generated by REDUCE-LAYOUT and for each schedule dimension either adds a reduce-vec expression to the circuit, if the dimension is exploded, or generates a rotate-and-reduce pattern, if the dimension is vectorized.

## 5.2 Array Materialization

Array materializers allow the compiler to customize how a layout is applied to an array traversal. They can be triggered to run only for certain array traversals and layouts, and thus can use specialized information about these to enable complex translations.

Array materializers implement two main functions. The MATERIALIZE-INPUT function materializes an array traversal indexing an input array. It takes as input the shape of the indexed array ($sh_a$), the array traversal itself ($at$), and the layout for the traversal specified by the schedule ($\ell$). The MATERIALIZE-EXPR function materializes an array traversal indexing an array that is the output of a let-bound statement. It takes similar input to MATERIALIZE-INPUT with the addition of the output layout of the indexed array ($o\ell_a$).

***Vector Derivation.*** The prototype implementation of the Viaduct-HE compiler has two array materializers. The first is the default materializer that is triggered on layouts with no preprocessing. When materializing traversals of input arrays, it attempts to minimize the number of input vectors required by *deriving* vectors from one another. When materializing traversals of let-bound arrays, it attempts to derive vectors of the traversal from the vectors defined by the output layout of the indexed array; materialization fails if some vector for the traversal cannot be derived.

Intuitively, a vector $v_1$ can be derived from another vector $v_2$ if all the array elements traversed by $v_1$ are contained in $v_2$ in the same relative positions, although rotation and masking might be required for the derivation. For example, consider the layout $\ell$ for traversal kt of 4x4 client input array k:

$$\text{kt} = \text{k}(0,0)[(2, \{0 :: 1\}), (4, \{0 :: 1\})] \qquad \ell = \{(i)\ 0 : 2 :: 1\}[1 : 4 :: 1].$$

Applying $\ell$ to kt yields two vectors:

$$\{i \mapsto 0\} \mapsto \text{k}(1,0)[(3,0,1,\{0::1\})] \qquad \{i \mapsto 1\} \mapsto \text{k}(0,0)[(4,0,0,\{0::1\})].$$

Then the vector at $\{i \mapsto 0\}$ can be derived from the vector at $\{i \mapsto 1\}$ by rotating the latter by -1 and masking its 4th slot. The materializer then generates the circuit expression $\textbf{rot}(\varphi_o, \varphi_c) \times \varphi_p$ for

$$ce + 0 = ce \qquad ce \times 1 = ce \qquad ce \times 0 = 0 \qquad ce_1 \times (ce_2 + ce_3) = (ce_1 \times ce_2) + (ce_1 \times ce_3)$$

$$\textbf{rot}(oe_1 + oe_2, ce) = \textbf{rot}(oe_1, \textbf{rot}(oe_2, ce)) \qquad \textbf{rot}(oe, ce_1) + \textbf{rot}(oe, ce_2) = \textbf{rot}(oe, ce_1 + ce_2)$$

$$\textbf{reduce-vec}_+(d : n, z \times ce) = z \times \textbf{reduce-vec}_+(d : n, ce)$$

$$\frac{d \notin \text{dim-vars}(oe)}{\textbf{reduce-vec}_\odot(d : n, \textbf{rot}(oe, ce)) = \textbf{rot}(oe, \textbf{reduce-vec}_\odot(d : n, ce))}$$

Fig. 8. Select identities for circuit optimization.

the kt and adds the following mappings to the registry:

$\varphi_c \mapsto \{\{i \mapsto 0\} \mapsto \text{Vec}(k(0,0)[(4,0,0,\{0 :: 1\})]), \{i \mapsto 1\} \mapsto \text{Vec}(k(0,0)[(4,0,0,\{0 :: 1\})])\}$

$\varphi_o \mapsto \{\{i \mapsto 0\} \mapsto -1, \{i \mapsto 1\} \mapsto 0\}$

$\varphi_p \mapsto \{\{i \mapsto 0\} \mapsto \text{Mask}((4,0,2)), \{i \mapsto 1\} \mapsto \text{Const}(1)\}$.

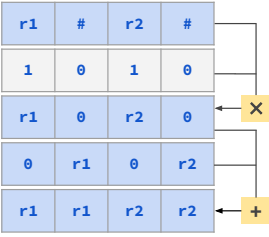

Besides rotation and masking, if a vector has an empty dimension it can be derived from a vector that contains a reduced dimension in the same position using a "clean-and-fill" routine, seen on the left [Aharoni et al. 2023][Dathathri et al. 2019, Figure 1]. This is useful for deriving vectors of traversals that index let-bound arrays.

***Roll Materializer.*** The other array materializer used by the Viaduct-HE compiler is specifically for layouts with a roll preprocessing operation. Given preprocessing $\text{roll}(a, b)$, the materializer splits on three cases, depending on the contents of traversal dimensions $a$ and $b$. In the first case, the roll materializer acts like the default materializer. In the second case, the materializer generates generalized diagonal vectors, similar to how the vectors of traversal at are generated in Figure 5b. In the third case, the materializer generates a subset of the vectors like the default materializer and then generates the rest by rotating these vectors, similar to how the vectors of xt are generated in the same figure. We defer to the supplementary materials for more details.

## 6 Circuit Transformations

Once a circuit is generated for the source program, the compiler has additional stages to further optimize the circuit before it generates target code.

### 6.1 Circuit Optimization

The scheduling stage of the compiler can find schedules with data layouts that result in efficient HE programs. However, there are optimizations leveraging the algebraic properties of HE operations that are missed by scheduling. The circuit optimization stage uses these algebraic properties to rewrite the circuit into an equivalent but more efficient form. The compiler performs efficient term rewriting through equality saturation [Tate et al. 2009; Willsey et al. 2021], applying rewrites to an e-graph data structure that compactly represents many equivalent circuits.

$$\mathbf{C}_e(\varphi_c, m) = (0, \mathbf{cipher}) \qquad \mathbf{C}_e(\varphi_p, m) = (0, \mathbf{plain})$$

$$\frac{\mathbf{C}_e(ce_1, m) = (v_1, \tau_1) \quad \mathbf{C}_e(ce_2, m) = (v_2, \tau_2)}{\mathbf{C}_e(ce_1 \odot ce_2, m) = (v_1 + v_2 + m\mathbf{W}(\odot_{\tau_1, \tau_2}), \tau_1 \sqcup \tau_2)}$$

$$\frac{\mathbf{C}_e(ce, mn) = (v, \tau)}{\mathbf{C}_e(\mathbf{reduce\text{-}vec}_\odot(d:n, ce), m) = (v + m(n-1)\mathbf{W}(\odot_{\tau, \tau}), \tau)}$$

$$\frac{\mathbf{C}_e(ce, m) = (v, \tau)}{\mathbf{C}_e(\mathbf{rot}(oe, ce), m) = (v + m\mathbf{W}(\mathrm{rot}_\tau), \tau)} \qquad \frac{\mathbf{C}_e(ce, \Pi_i\, n_i) = (v_{ce}, \tau) \quad \mathbf{C}_s(cs) = v_{cs}}{\mathbf{C}_s(\mathbf{let}\ a : [d_1:n_1, \ldots, d_n:n_n] = ce\ \mathbf{in}\ cs) = v_{ce} + v_{cs}}$$

Fig. 9. Circuit cost function.

Figure 8 contains some identities that hold for circuit expressions. Because homomorphic addition and multiplication operate element-wise, one can view HE programs algebraically as product rings; thus the usual ring properties hold. Circuit identities also express properties of rotations and reductions. For example, rotation distributes over addition and multiplication: adding or multiplying vectors and then rotating yields the same result as rotating the vectors individually first and then adding or multiplying. Provided that the rotation amount $oe$ does not depend on the value of the dimension variable that is being reduced (i.e. the variable is not in dim-vars($oe$)) rotating vectors individually by $oe$ and then reducing them together is the same as reducing the vectors first and then rotating the result.

*Computing cost.* Extraction of efficient circuits from the e-graph is guided by the cost function defined in Figure 9. Note that this is the same cost function that guides the search for efficient schedules during the scheduling stage. The function $\mathbf{C}_e$ takes an expression $ce$ and its multiplicity $m$ and returns the cost $v$ of the expression as well as its type $\tau$, which could either be **plain** or **cipher**. Types are ordered such that **plain** $\sqsubseteq$ **cipher**; the type for binary operations is computed from the join of its operand types according to this ordering. The cost function is parameterized by a customizable function $\mathbf{W}$ that weights operations according to their type. For example, $\mathbf{W}$ might give greater cost to operations between ciphertexts than to operations between plaintexts. The cost function also adds costs for other features of a circuit, such as the number of input vectors required (which can be computed from the circuit registry) and the multiplication depth of circuits, which is an important proxy metric for ciphertext noise that should be minimized to avoid needing using costlier encryption parameters [Cowan et al. 2021].

## 6.2 Plaintext Hoisting

Not all data in an HE program are ciphertexts; instead some data such as constants and server inputs are plaintexts. Because plaintext values are known by the server, operations between such values can be executed natively, which is more efficient than execution under HE. The plaintext hoisting stage finds circuit components that can be hoisted out and executed natively.

The compiler performs plaintext hoisting by finding maximal circuit subexpressions that perform computations only on plaintexts. Once a candidate subexpression is found, the compiler creates a let statement with the subexpression as its body. In the original circuit, the subexpression is replaced with a plaintext variable; in the circuit registry this variable is mapped to vectors that reference the output of the created let statement.

| | | |
|---|---|---|
| Instruction ID $i$ | Array name $a$ | Dimension name $d$   Vector $v$ |
| Value type | $\tau_v$ ::= | **native** (**N**) \| **plain** (**P**) \| **cipher** (**C**) |
| Instruction type | $\tau_i$ ::= | **native** (**N**) \| **cipher-plain** (**CP**) \| **cipher** (**C**) |
| Array reference | $\rho_a$ ::= | $a \mid \rho_a[d]$ |
| Reference | $\rho$ ::= | $i \mid \rho_a$ |
| Constructor | $c$ ::= | $\text{Array}_{\tau_v}(\overline{n}) \mid \text{Const}(n) \mid \text{Mask}(\overline{(n,n,n)}) \mid \text{Vec}(v)$ |
| Expression | $le$ ::= | $z \mid le \odot le \mid d \mid \rho \mid c \mid \textbf{encode}(\rho_a)$ |
| Statement | $ls$ ::= | **skip** $\mid ls; ls \mid i \leftarrow \odot_{\tau_i}(\rho, \rho) \mid i \leftarrow \text{rot}_{\tau_i}(le, \rho)$ |
| | $\mid$ | $\rho_a := le \mid \textbf{for } d \textbf{ in range}(n) \{ ls \}$ |

Fig. 10. Abstract syntax for loop-nest programs.

## 7 Circuit Lowering

The circuit representation facilitates optimizations but is hard to translate into target code. The circuit lowering stage takes a circuit program as input and generates a *loop-nest* program that closely resembles target code. Back ends then only need to translate loop-nest programs to target code to add compiler support for HE libraries.

Figure 10 defines the abstract syntax of loop-nest programs. Programs manipulate arrays of vectors, which can come in three different value types. Native vectors represent data in the "native" machine representation; they cannot be used in HE computations. Plaintext vectors are encoded as HE plaintexts and can be used in HE computations. Ciphertext vectors are encrypted data from the client. Computations are represented as sequences of instructions, which are tagged with an instruction type that represents the types of their operands. Statements include instructions, assignments to arrays, and for loops. Server inputs are first declared as native vectors, and then explicitly encoded into plaintexts using the **encode** expression. Explicit representation of encoding allows the compiler to generate code to encode the results of computations over native vectors.

Circuit lowering translates a circuit statement **let** $a : [d_1 : n_1, \ldots, d_n : n_n] = ce$ by first generating a sequence of prelude statements that fill arrays with registry values for offset, ciphertext, and plaintext variables used in $ce$. The translation for the statement itself consists of a nest of $n$ for-loops, one for each of the dimension variables $d_i$. The body of the loop nest is the translation for $ce$.

Translation of most expression forms are straightforward. Literals are replaced with references to plaintext vectors that contain the literal value in all slots. Operations and rotations are translated as instructions. The expression form **reduce-vec**$_\odot(d : n, ce)$ is translated by declaring an array $a$ and a new loop that iterates over $d$. The body of the loop contains the translation for $ce$ and its output is stored in the newly declared array $a$. After the loop, a sequence of instructions then computes the reduction as a balanced tree of operations; this is particularly important to minimize multiplication depth.[3]

Finally, a value numbering analysis over circuits prevents redundant computations in the translation to loop-nest instructions. For example, in Figure 3c the difference between a test point and the client-provided point is only computed once; the result is then multiplied with itself to compute the squared difference.

## 8 Implementation

We have implemented a prototype version of the Viaduct-HE compiler in about 13k LoC of Rust. The compiler uses the egg [Willsey et al. 2021] equality saturation library for the circuit optimization stage. We configure egg to use the LP extractor, which lowers e-graph extraction as an integer linear program.[4] The compiler's cost estimator is tuned to reflect the relative latencies of operations and to give lower cost to plaintext-plaintext operations than ciphertext-ciphertext or ciphertext-plaintext operations (which must be executed in HE), driving the optimization stage toward circuits with plaintext hoistable components.

---

[3]When reducing with addition, where noise growth is not a concern, the loop instead accumulates values directly into $a$ at the end of each iteration.

| Benchmark | Vector Size | Configuration Exec Time (s) | | | |
|---|---|---|---|---|---|
| | | baseline | e1-o0 | e2-o0 | e2-o1 |
| conv-simo | 4096 | 62.21 | 0.10 | — | **0.09** |
| conv-siso | 4096 | 15.58 | 0.04 | — | **0.03** |
| distance | 2048 | 0.54 | 0.37 | **0.17** | — |
| double-matmul | 4096 | 74.84 | **0.07** | — | — |
| retrieval-256 | 8192 | 120.12 | **0.70** | — | — |
| retrieval-1024 | 8192 | 585.08 | 1.92 | **1.01** | — |
| set-union-16 | 8192 | 93.98 | **1.01** | — | — |
| set-union-128 | 16384 | >3600 | **11.65** | — | — |

Fig. 11. Execution time for benchmark configurations, in seconds.

| Benchmark | Scheduling (s) | | Circuit Opt (s) |
|---|---|---|---|
| | e1 | e2 | o1 |
| conv-simo | 9.43 | 100.03 | 0.003 |
| conv-siso | 1.27 | 14.09 | 0.08 |
| distance | 0.04 | 6.28 | 6.42 |
| double-matmul | 0.64 | 5.84 | 42.47 |
| retrieval-256 | 0.05 | 0.80 | 170.45 |
| retrieval-1024 | 0.56 | 16.99 | 5.52 |
| set-union-16 | 0.06 | 1.66 | 3.60 |
| set-union-128 | 7.59 | 663.23 | 9.71 |

Fig. 12. Compilation time for benchmark configurations, in seconds.

We have implemented a back end that targets the BFV [Fan and Vercauteren 2012] scheme implementation of the SEAL homomorphic encryption library [Chen et al. 2017]. The compiler generates Python code that calls into SEAL using the PySEAL [Titus et al. 2018] library. The back end consists of about 1k LoC of Rust and an additional 500 lines of Python. It performs a use analysis to determine when memory-efficient in-place versions of SEAL operations can be used. We use the numpy library [Harris et al. 2020] to pack arrays into vectors.

## 9 Evaluation

To evaluate Viaduct-HE, we ran experiments to determine the efficiency of vectorized HE programs generated by the compiler and to determine whether its compilation process is scalable. We used benchmarks that are either common in the literature or have been adapted from prior work. Our benchmarks are larger than those used to evaluate Porcupine [Cowan et al. 2021] and Coyote [Malik et al. 2023].

---

[4]The default LP extractor implementation of the egg library uses the COIN-OR CBC solver [COIN-OR 4 11].

*Experimental setup.* We ran experiments on a Dell OptiPlex 7050 machine with an 8-core Intel Core i7 7th Gen CPU and 32 GB of RAM. All numbers reported are averaged over 5 trials, with relative standard error below 8 percent.[5] We use the following programs as benchmarks:

- **conv**. A convolution over a 1-channel 32x32 client-provided image with a server-provided filter of size 3 and stride 1. The **conv-siso** variant (**s**ingle-**i**nput, **s**ingle-**o**utput) applies a single filter to the image, while the **conv-simo** variant (**s**ingle-**i**nput, **m**ultiple-**o**utput) applies 4 filters to the image.
- **distance-64**. The distance program from §3, but points have 64 dimensions and there are 64 test points.
- **double-matmul**. Given 16x16 matrices $A_1$, $A_2$, and $B$, computes $A_2 \times (A_1 \times B)$.
- **retrieval**. A private information retrieval example where the user queries a key-value store. The **retrieval-256** variant has 256 key-value pairs and 8 bit keys, while **retrieval-1024** has 1024 pairs and 10 bit keys.
- **set-union** (from Viand et al. [2022]). An aggregation from two key-value stores $A$ and $B$. The program sums all the values in $A$ and the values in $B$ that do not share a key with some value in $A$. In the **set-union-16** variant $A$ and $B$ each have 16 key-value pairs and 4 bit keys, while in **set-union-128** $A$ and $B$ each have 128 key-value pairs and 7 bit keys.

We compiled these programs with various target vector sizes, shown in Figure 11.[6] The source code and compiled programs for all benchmarks are in the supplementary materials.

## 9.1 Efficiency of Compiled Programs

To determine whether the Viaduct-HE compiler can generate efficient vectorized HE programs, we compared compiled benchmarks against baseline HE implementations using simple vectorization schedules. These baselines do not match the efficiency of expert-written implementations, but they illustrate the importance of vectorization schedules in the performance of HE programs. The baseline implementations are as follows:

- For **conv-siso** and **conv-simo**, each vector contains all the input pixels used to compute the value of a single output pixel.
- For **distance-64** the baseline implementation is the row-wise layout from Figure 5a.
- For **double-matmul** the input matrices $A_1$ and $B$ for the first multiplication are stored in vectors column- or row-wise to allow a single multiplication and then rotate-and-reduce to compute a single output entry. This output layout forces $A_2$ to be stored as one matrix entry per vector.
- For **retrieval** and **set-union**, keys and values are stored in individual vectors.

We compared baseline implementations against implementations generated with different configurations of the Viaduct-HE compiler. For scheduling and circuit optimization, we test two configurations each: **e1** schedules for one epoch, such that the tiling transformer is disabled; **e2** schedules for two epochs; **o0** disables circuit optimization; **o1** runs circuit optimization such that equality saturation stops after either a timeout of 60 seconds or an e-graph size limit of 500 e-nodes. We did not find any optimization improvements in further increasing these limits. We use the configuration combinations **e1-o0**, **e2-o0**, and **e2-o1** in experiments.

Figure 11 shows the results the average execution time of each benchmark under different configurations. We timed out the execution of the **set-union-128** baseline after 1 hour. For all benchmarks, Viaduct-HE implementations run faster than the baselines, with speedups ranging from 50 percent (1.45x for **distance-64** with configuration **e1-o0**) to several orders of magnitude (over 1000x for **double-matmul**). The bulk of the speedups come from the scheduling stage: only the **conv** variants show performance differences between **o0** and **o1**, since in most benchmarks circuit optimization generates the same initial circuit. We believe this is because the compiler already uses domain-specific techniques like rotate-and-reduce to generate efficient circuits before optimization, making it hard to improve on the initial circuit. Also note that most benchmarks found the optimal schedule after 1 epoch; only **distance** and **retrieval-1024** have more efficient schedules in configuration **e2** compared to **e1**.

---

[5]With the exception for the execution time reported for circuit optimization; there relative standard error is below 25 percent. The higher error is from the extractor calling into an external LP solver.
[6]The reported vector size in Figure 11 is half of the polynomial modulus degree parameter $N$, since in BFV vector slots are arranged as a $2 \times N/2$ matrix such that rotation cyclically shifts elements within rows.

| x1 | x3 | x5 | … | x63 | x1 | x3 | x5 | … |
| x2 | x4 | x6 | … | x64 | x2 | x4 | x6 | … |

Fig. 13. Layout for client point in **e2-o0** implementation of **distance-64**.

The implementations generated by Viaduct-HE make efficient use of the SIMD capabilities of HE with sophisticated layouts. For **distance-64**, the **e1-o0** configuration generates the diagonal layout from Figure 5b, which reduces the necessary amount of rotations and additions compared to the row-wise baseline layout. The **e2-o0** configuration generates an even more efficient layout by using all 2048 vector slots available: the even and odd coordinates of the client point are packed in separate vectors and each coordinate is repeated 64 times, allowing the squared difference of each even (resp. odd) coordinate with the corresponding even (resp. odd) coordinate of each test point to be computed simultaneously. Figure 13 shows the layout for the client point.

Meanwhile, for **retrieval-256** the compiler generates a layout where the entire key array and the query are each stored in single vectors. Each bit of the query is repeated 256 times, allowing the equality computation with the corresponding bit of each key to be computed all at once. For **retrieval-1024**, a similar layout to **retrieval-256** is not possible because there are too many keys to store in a single vector. Instead, the **e1** configuration explodes the key array bit-wise: each bit of a key is stored in a separate vector, and the corresponding bits of all 1024 keys are packed in the same vector. The **e2** configuration, as in **distance-64**, stores the even and odd bits of keys in separate vectors to use more of the available 8192 vector slots, making it even more efficient.

## 9.2 Comparison with Expert-written HE Programs

The HE programs generated by the Viaduct-HE compiler are not only dramatically more efficient than the baseline implementations, they also sometimes match or even improve upon expert-written implementations found in the literature.

The **conv-simo** implementation generated by Viaduct-HE is basically the "packed" convolution kernel defined in Gazelle [Juvekar et al. 2018]: both store the image in a single ciphertext, while the values of all 4 filters at a single position are packed in a single plaintext. The image ciphertext is then rotated to align with the filter ciphertexts; since the filter size is 3x3, 9 rotated image ciphertexts and 9 filter plaintexts are multiplied together, and then summed. This computes the convolution for all output pixels at once. The **conv-siso** implementation is similar, but instead packs *columns* of the filter into a plaintext instead of single values.

The **o1** configurations of **conv-siso** and **conv-simo** use algebraic properties of circuits to optimize the implementation further. Given image ciphertext $c$, mask $m$, and plaintext filter $f$, instead of computing $(c \times m) \times f$ in HE as two ciphertext-plaintext operations, circuit optimization rewrites the computation as $c \times (m \times f)$, allowing $m \times f$ to be hoisted out of HE and computed natively. This is exactly the "punctured plaintexts" technique, also from Gazelle.

For **double-matmul**, Viaduct-HE generates an implementation where each matrix is laid out in a single vector. Importantly, even though $A_1$ and $A_2$ are both left operands to multiplication, their layouts are different because the layout of $A_2$ must account for the output layout of $A_1 \times B$. The generated implementation is similar to the expert implementation found in Dathathri et al. [2019, Figure 1], but avoids a "clean-and-fill" operation required to derive an empty dimension from a reduced vectorized dimension. The Viaduct-HE implementation avoids this operation by moving the reduced vectorized dimension as the outermost dimension in the vector, thus making it a reduced repeated dimension in the output layout of the first multiplication. The expert implementation takes 0.06 seconds compared to 0.04 seconds taken by the Viaduct-HE implementation, a 1.5x speedup.

Finally, **set-union-128** is originally a benchmark for the HECO compiler [Viand et al. 2022]. The program computes a mask that zeroes out elements of $B$ with keys that are in $A$, and then adds the sum of values in $A$ with the sum of masked values in $B$. The implementation generated by the HECO compiler is over 40x slower than an expert-written solution: as in the **e1** configuration of **retrieval-1024**, it packs each bit of a key in separate vector, and the corresponding bits of all 128 keys are packed in the same vector. The bits are repeated within each vector such that the computation of masks for all pairs of keys in $A$ and $B$ can be done simultaneously. The Viaduct-HE compiler generates exactly this expert solution.

### 9.3 Scalability of Compilation

To determine whether the Viaduct-HE compilation process is scalable, we measured the compilation time for each benchmark, with the same scheduling and optimization configurations from RQ1. Figure 12 shows the compilation times for the benchmarks. The two main bottlenecks for compilation are the scheduling and circuit optimization stages, with their sum constituting almost all compilation time.

With 1 epoch, scheduling at most takes 10 seconds; however, with 2 epochs scheduling takes up to 11 minutes (**set-union-128**). This is because the tiling transformer vastly increases the search space, as it finds many different ways to split dimensions. We find that scheduling is mainly hampered by the fact that circuit generation must be attempted for every visited schedule, as it is currently the only way to determine whether a schedule is valid. In particular, circuit generation is greatly slowed down by array materialization, as in many schedules (especially those with many exploded dimensions), the default array materializer generates thousands of vectors and then tries to derive these vectors from one another, so that scheduler has an accurate count of features such as the number of input vectors and rotations. Speeding up scheduling by estimating such features without array materialization is an interesting research direction.

Meanwhile, circuit optimization time is completely dominated by extraction. In all compilations, equality saturation stops in less than a second, but extraction takes longer (almost 3 minutes for **retrieval-256**) because the LP extractor must solve an integer linear program.

## 10 Related Work

*Vectorized HE for Specific Applications.* There is a large literature on developing efficient vectorized HE implementations of specific applications, particularly for machine learning. Some work such as Cryptonets [Gilad-Bachrach et al. 2016], Gazelle [Juvekar et al. 2018], LoLa [Brutzkus et al. 2019], and HyPHEN [Kim et al. 2023] develop efficient vectorized kernels for neural network inference. Other work such as SEALion [van Elsloo et al. 2019] and nGraph-HE [Boemer et al. 2019a,b] provide domain-specific compilers for neural networks. CHET [Dathathri et al. 2019] automatically selects from a fixed set of data layouts for neural network inference kernels. HeLayers [Aharoni et al. 2023] is similar to CHET in automating layout selection, but can also search for efficient tiling sizes for kernels, akin to the tiling transformer in Viaduct-HE. COPSE [Malik et al. 2021] develops a vectorized implementation of decision forest evaluation.

*Compilers for HE.* The programmability challenges of HE have inspired much recent work on HE compilers [Viand et al. 2021]. HE compilers face similar challenges as compilers for multi-party computation [Acay et al. 2021; Büscher et al. 2018; Chen et al. 2023; Hastings et al. 2019; Rastogi et al. 2014], such as lowering programs to a circuit representation. At the same time, HE has unique programmability challenges that are not comparable to other domains. Some HE compilers such as Alchemy [Crockett et al. 2018], Cingulata [Carpov et al. 2015], EVA [Dathathri et al. 2020], HECATE [Lee et al. 2022], and Ramparts [Archer et al. 2019], focus on other programmability concerns besides vectorization, such as selection of encryption parameters and scheduling "ciphertext maintenance" operations. Lobster [Lee et al. 2020] uses program synthesis and term rewriting to optimize HE circuits, but it focuses on boolean circuits and not on vectorized arithmetic circuits.

Recent work have tackled the challenge of automatically vectorizing programs for HE. Porcupine [Cowan et al. 2021] proposes a synthesis-based approach to generating vectorized HE programs from an imperative source program. However, Porcupine requires the developer to provide the data layout for inputs and can only scale up to HE programs with a small number of instructions. HECO [Viand et al. 2022] attempts to solve the scalability issue by analyzing indexing operations in the source program in lieu of program synthesis, but fixes a simple layout for all programs, leaving many optimization opportunities out of reach. Coyote [Malik et al. 2023] uses search and LP to find efficient vectorizations of arithmetic circuits, balancing vectorization opportunities with data movement costs. Coyote can vectorize "irregular" programs that are out of scope for Viaduct-HE. At the same time, though it can generate layouts for HE programs, Coyote still requires user hints for "noncanonical" layouts. Also, Coyote appears to be less scalable than Viaduct-HE, as compiling a 16x16 matrix multiplication requires decomposition into 4x4 matrices that are "blocked" together.

*Array-oriented Languages.* In the taxonomy given by Paszke et al. [2021], the Viaduct-HE source language is a "pointful" array-oriented language with explicit indexing constructs, in contrast to array "combinator" languages such as Futhark [Henriksen et al. 2017] and Lift [Steuwer et al. 2017]. The Viaduct-HE source language is thus similar in spirit to languages such as ATL [Bernstein et al. 2020; Liu et al. 2022], Dex [Paszke et al. 2021], and Tensor Comprehensions [Vasilache et al. 2018]. In particular, the separation of algorithm and schedule in Viaduct-HE is inspired by the Halide [Ragan-Kelley et al. 2013] language and compiler for image processing pipelines. Although the source language of Viaduct-HE is similar to Halide's—both are pointful array languages—Viaduct-HE schedules have very different concerns from Halide schedules. On one hand, Halide schedules represent choices such as what order the values of an image processing stage should be computed, and the granularity at which stage results are stored; on the other hand, Viaduct-HE schedules represent the layout of data in ciphertext and plaintext vectors.

## 11 Summary

With its array-oriented source language, the Viaduct-HE compiler can give a simple representation for vectorization schedules and find sophisticated data layouts comparable to expert HE implementations. The compiler also has representations to allow for algebraic optimizations and for easy implementation of back ends for new HE libraries. Overall, the Viaduct-HE compiler drastically lowers the programmability burden of vectorized homomorphic encryption.

## A  Roll Preprocessing

The roll preprocessor "rolls" the elements of traversal dimension $a$ progressively along traversal dimension $b$. That is, given that $a$ and $b$ have the same extent $n$,

$$\text{roll}(a,b) = \lambda(x_1, \ldots, x_a, \ldots, x_b, \ldots, x_n).(x_1, \ldots, x_a + x_b \,\%\, n, \ldots, x_b, \ldots x_n).$$

### A.1  Roll Transformer

The roll transformer applies a roll preprocessing operation to a layout:

$$\text{id}\{\ldots, (d_a)\ i_a : n :: s_a, \ldots\}[i_b : n :: s_b, \ldots] \rightsquigarrow \text{roll}(a,b)\{\ldots, (d_a)\ i_a : n :: s_a, \ldots\}[i_b : n :: s_b, \ldots].$$

The transformer only applies when dimension $a$ is exploded, dimension $b$ is the outermost vectorized dimension and their extents match. The following conditions must also hold:

- the traversal dimensions $a$ and $b$ are not tiled;
- $a$ and $b$ have content dimension sets with size at most 1;
- if $a$ or $b$ has content dimension set $\{d :: s\}$, the extent of $d$ must be the same as the extents of $a$ and $b$ and stride $s$ must equal 1;
- the dimension in the indexed array traversed by $a$ and $b$, if any, must not be traversed in other dimensions.

These conditions ensure that layouts with roll preprocessing can be materialized.

### A.2  Roll Array Materializer

The roll array materializer generates vectors for an array traversal with a layout containing preprocessing $\text{roll}(a, b)$. Given traversal dimensions $a$ and $b$, let $a = (n, cd_a)$ and $b = (n, cd_b)$. There are three possible cases:

- **Case 1**: $cd_a = \{\}$. Then the contents of the array traversal do not change along $a$, so the roll preprocessing does not do anything. The traversal will be materialized as if the layout has no preprocessing.
- **Case 2**: $cd_a = \{i_a :: 1\}$ and $cd_b = \{i_b :: 1\}$. Then the traversal will be materialized as if the layout has no preprocessing, but the vectors generated will have preprocessing $\text{roll}(i_a, i_b)$.
- **Case 3**: $cd_a = \{i_a :: 1\}$ and $cd_b = \{\}$. Then the traversal will be materialized as if the layout has no preprocessing, but only for vectors where $\{b \mapsto 0\}$. To materialize a vector at coordinate $c = \{\ldots, b \mapsto v, \ldots\}$ where $v \neq 0$, the vector at coordinate $c[b \mapsto 0]$ (i.e. $c$ but with $b$ set to 0) is rotated amount $v$.

***Example.*** Consider the distance example from Figure 3 compiled with the diagonal layout. Traversals $at$ and $xt$ are defined as follows:

$$at = a(0,0)[(4, \{0 :: 1\}), (4, \{1 :: 1\})] \qquad xt = x(0)[(4, \{\}), (4, \{0 :: 1\})].$$

The diagonal layout is defined as $\text{roll}(1,0)\{(i) \; 1 : 4 :: 1\}[0 : 4 :: 1]$. Thus when we apply the layout to $at$, case 2 above holds. So the materializer returns ciphertext variable $\varphi_c$ as the circuit expression representing $at$ and $\varphi_c$ is mapped to the following in the circuit registry:

$$\{i \mapsto 0\} \mapsto \text{a.roll}(1,0)(0,0)[(4,0,0,\{0::1\})] \qquad \{i \mapsto 1\} \mapsto \text{a.roll}(1,0)(0,1)[(4,0,0,\{0::1\})]$$
$$\{i \mapsto 2\} \mapsto \text{a.roll}(1,0)(0,2)[(4,0,0,\{0::1\})] \qquad \{i \mapsto 3\} \mapsto \text{a.roll}(1,0)(0,3)[(4,0,0,\{0::1\})].$$

When we apply the layout to $xt$, case 3 holds. Thus the materializer returns $\mathbf{rot}(i, \varphi_c)$ as the circuit expression representing $xt$ and $\varphi_c$ is mapped to the following in the circuit registry:

$$\{i \mapsto 0\} \mapsto \text{x}(0)[(4,0,0,\{0::1\})] \qquad \{i \mapsto 1\} \mapsto \text{x}(0)[(4,0,0,\{0::1\})]$$
$$\{i \mapsto 2\} \mapsto \text{x}(0)[(4,0,0,\{0::1\})] \qquad \{i \mapsto 3\} \mapsto \text{x}(0)[(4,0,0,\{0::1\})].$$

Note that an offset variable is not needed here, because the value of the rotation coincides exactly with the value that the exploded dimension $i$ takes.

## B Benchmarks

This section contains the source code and implementations generated by the Viaduct-HE compiler, given in the loop-nest representation.

### B.1 Source code

**conv-simo**

```
input img: [32,32] from client
input filter: [4,3,3] from server
for x: 30 {
  for y: 30 {
    for out: 4 {
      sum(for i: 3 {
        sum(for j: 3 {
          img[x + i][y + j] * filter[out][i][j]
        })
      })
    }
  }
}
```
**conv-siso**

```
input img: [32,32] from client
input filter: [3,3] from server
for x: 30 {
  for y: 30 {
    sum(for i: 3 {
      sum(for j: 3 {
        img[x + i][y + j] * filter[out][i][j]
      })
    })
  }
}
```

**distance**

```
input point: [64] from client
input tests: [64,64] from server
for i: 64 {
  sum(for j: 64 {
    (point[j] - tests[i][j]) * (point[j] - tests[i][j])
  })
}
```

**matmul-2**

```
input A1: [16,16] from server
input A2: [16,16] from server
input B: [16,16] from client
let res =
  for i: 16 {
    for j: 16 {
      sum(for k: 16 { A1[i][k] * B[k][j] })
    }
  }
in
for i: 16 {
  for j: 16 {
    sum(for k: 16 { A2[i][k] * res[k][j] })
  }
}
```

**retrieval-256**

```
input keys: [256,8] from client
input values: [256] from client
input query: [8] from client
let mask =
  for i: 256 {
    product(for j: 8 {
      1 - ((query[j] - keys[i][j]) * (query[j] - keys[i][j]))
    })
  }
in
```

```
sum(values * mask)
```

**retrieval-1024**

```
input keys: [1024,10] from client
input values: [1024] from client
input query: [10] from client
let mask =
  for i: 1024 {
    product(for j: 10 {
      1 - ((query[j] - keys[i][j]) * (query[j] - keys[i][j]))
    })
  }
in
sum(values * mask)
```

**set-union-16**

```
input a_id: [16, 4] from client
input a_data: [16] from client
input b_id: [16, 4] from client
input b_data: [16] from client
let a_sum = sum(a_data) in
let b_sum =
  sum(for j: 16 {
    b_data[j] *
    product(for i: 16 {
      1 -
      product(for k: 4 {
        1 - ((a_id[i][k] - b_id[j][k]) * (a_id[i][k] - b_id[j][k]))
      })
    })
  })
in
a_sum + b_sum
```

**set-union-128**

```
input a_id: [128, 7] from client
input a_data: [128] from client
input b_id: [128, 7] from client
input b_data: [128] from client
let a_sum = sum(a_data) in
let b_sum =
  sum(for j: 128 {
    b_data[j] *
    product(for i: 128 {
      1 -
      product(for k: 7 {
        1 - ((a_id[i][k] - b_id[j][k]) * (a_id[i][k] - b_id[j][k]))
      })
    })
```

```
  })
in
a_sum + b_sum
```

## B.2 Implementations
### conv-simo e1-o0

```
val v_img_1: C = vector(img(0, 0)[(32, 0, 0 {1 :: 1}), (4, 0, 0, {}),
    (32, 0, 0 {0 :: 1})])
val v_filter_1: N = vector(filter(0, 2, 2)[(30, 0, 2, {}), (4, 0, 0 {0 ::
    1}), (30, 0, 2, {})])
val v_filter_2: N = vector(filter(0, 2, 1)[(30, 0, 2, {}), (4, 0, 0 {0 ::
    1}), (30, 0, 2, {})])
val v_filter_3: N = vector(filter(0, 0, 1)[(30, 0, 2, {}), (4, 0, 0 {0 ::
    1}), (30, 0, 2, {})])
val v_filter_4: N = vector(filter(0, 1, 2)[(30, 0, 2, {}), (4, 0, 0 {0 ::
    1}), (30, 0, 2, {})])
val v_filter_5: N = vector(filter(0, 2, 0)[(30, 0, 2, {}), (4, 0, 0 {0 ::
    1}), (30, 0, 2, {})])
val v_filter_6: N = vector(filter(0, 0, 2)[(30, 0, 2, {}), (4, 0, 0 {0 ::
    1}), (30, 0, 2, {})])
val v_filter_7: N = vector(filter(0, 0, 0)[(30, 0, 2, {}), (4, 0, 0 {0 ::
    1}), (30, 0, 2, {})])
val v_filter_8: N = vector(filter(0, 1, 0)[(30, 0, 2, {}), (4, 0, 0 {0 ::
    1}), (30, 0, 2, {})])
val v_filter_9: N = vector(filter(0, 1, 1)[(30, 0, 2, {}), (4, 0, 0 {0 ::
    1}), (30, 0, 2, {})])
val mask_1: N = mask([(32, 0, 30), (4, 0, 3), (32, 0, 30)])
encode(v_filter_3)
encode(v_filter_4)
encode(v_filter_1)
encode(v_filter_9)
encode(v_filter_5)
encode(v_filter_7)
encode(v_filter_8)
encode(v_filter_2)
encode(v_filter_6)
encode(mask_1)
var pt2: P[3][3] = 0
pt2[0][0] = v_filter_7
pt2[0][1] = v_filter_8
pt2[0][2] = v_filter_5
pt2[1][0] = v_filter_3
pt2[1][1] = v_filter_9
pt2[1][2] = v_filter_2
pt2[2][0] = v_filter_6
pt2[2][1] = v_filter_4
pt2[2][2] = v_filter_1
```

```
var __out: C = 0
var __reduce_2: C = 0
for i2 in range(3) {
    var __reduce_1: C = 0
    for i7 in range(3) {
        instr1 = rot(CC, ((0 + (-128 * i7)) + (-1 * i2)), v_img_1)
        instr3 = mul(CP, instr1, mask_1)
        instr5 = mul(CP, instr3, pt2[i7][i2])
        instr6 = add(CC, __reduce_1, instr5)
        __reduce_1 = instr6
    }
    instr8 = add(CC, __reduce_2, __reduce_1)
    __reduce_2 = instr8
}
__out = __reduce_2
```

**conv-simo e2-o1**

```
val v_img_1: C = vector(img(0, 0)[(32, 0, 0 {0 :: 1}), (4, 0, 0, {}),
    (32, 0, 0 {1 :: 1})])
val v_filter_1: N = vector(filter(0, 0, 2)[(30, 0, 2, {}), (4, 0, 0 {0 ::
    1}), (30, 0, 2, {})])
val v_filter_2: N = vector(filter(0, 2, 0)[(30, 0, 2, {}), (4, 0, 0 {0 ::
    1}), (30, 0, 2, {})])
val v_filter_3: N = vector(filter(0, 1, 1)[(30, 0, 2, {}), (4, 0, 0 {0 ::
    1}), (30, 0, 2, {})])
val v_filter_4: N = vector(filter(0, 0, 0)[(30, 0, 2, {}), (4, 0, 0 {0 ::
    1}), (30, 0, 2, {})])
val v_filter_5: N = vector(filter(0, 2, 1)[(30, 0, 2, {}), (4, 0, 0 {0 ::
    1}), (30, 0, 2, {})])
val v_filter_6: N = vector(filter(0, 0, 1)[(30, 0, 2, {}), (4, 0, 0 {0 ::
    1}), (30, 0, 2, {})])
val v_filter_7: N = vector(filter(0, 1, 2)[(30, 0, 2, {}), (4, 0, 0 {0 ::
    1}), (30, 0, 2, {})])
val v_filter_8: N = vector(filter(0, 2, 2)[(30, 0, 2, {}), (4, 0, 0 {0 ::
    1}), (30, 0, 2, {})])
val v_filter_9: N = vector(filter(0, 1, 0)[(30, 0, 2, {}), (4, 0, 0 {0 ::
    1}), (30, 0, 2, {})])
val mask_1: N = mask([(32, 0, 30), (4, 0, 3), (32, 0, 30)])
var pt2: P[3][3] = 0
pt2[0][0] = v_filter_4
pt2[0][1] = v_filter_9
pt2[0][2] = v_filter_2
pt2[1][0] = v_filter_6
pt2[1][1] = v_filter_3
pt2[1][2] = v_filter_5
pt2[2][0] = v_filter_1
pt2[2][1] = v_filter_7
pt2[2][2] = v_filter_8
```

```
var __partial_1: N[3][3] = 0
for i2 in range(3) {
    for i6 in range(3) {
        instr2 = mul(N, mask_1, pt2[i6][i2])
        __partial_1[i2][i6] = instr2
    }
}
for i2 in range(3) {
    for i6 in range(3) {
        encode(__partial_1[i2][i6])
    }
}
var __out: C = 0
var __reduce_2: C = 0
for i2 in range(3) {
    var __reduce_1: C = 0
    for i6 in range(3) {
        instr4 = rot(CC, ((-128 * i2) + (-1 * i6)), v_img_1)
        instr6 = mul(CP, instr4, __partial_1[i2][i6])
        instr7 = add(CC, __reduce_1, instr6)
        __reduce_1 = instr7
    }
    instr9 = add(CC, __reduce_2, __reduce_1)
    __reduce_2 = instr9
}
__out = __reduce_2
```

**conv-siso e1-o0**

```
val v_img_1: C = vector(img(0, 0)[(30, 0, 2 {0 :: 1}), (3, 0, 1 {0 :: 1})
    , (32, 0, 0 {1 :: 1})])
val v_filter_1: N = vector(filter(0, 0)[(30, 0, 2, {}), (3, 0, 1 {0 ::
    1}), (30, 0, 2, {})])
val v_filter_2: N = vector(filter(0, 1)[(30, 0, 2, {}), (3, 0, 1 {0 ::
    1}), (30, 0, 2, {})])
val v_filter_3: N = vector(filter(0, 2)[(30, 0, 2, {}), (3, 0, 1 {0 ::
    1}), (30, 0, 2, {})])
val mask_1: N = mask([(32, 0, 30), (4, 0, 3), (32, 0, 30)])
encode(v_filter_3)
encode(v_filter_2)
encode(v_filter_1)
encode(mask_1)
var pt2: P[3] = 0
pt2[0] = v_filter_1
pt2[1] = v_filter_2
pt2[2] = v_filter_3
var __out: C = 0
var __reduce_1: C = 0
for i4 in range(3) {
```

```
    instr1 = rot(CC, (0 + (-1 * i4)), v_img_1)
    instr3 = mul(CP, instr1, mask_1)
    instr5 = mul(CP, instr3, pt2[i4])
    instr6 = add(CC, __reduce_1, instr5)
    __reduce_1 = instr6
}
instr8 = rot(CC, -64, __reduce_1)
instr9 = add(CC, __reduce_1, instr8)
instr10 = rot(CC, -32, instr9)
instr11 = add(CC, instr9, instr10)
__out = instr11
```

**conv-siso e2-o1**

```
val v_img_1: C = vector(img(0, 0)[(3, 0, 1 {0 :: 1}), (30, 0, 2 {0 :: 1})
    , (32, 0, 0 {1 :: 1})])
val v_filter_1: N = vector(filter(0, 0)[(3, 0, 1 {0 :: 1}), (30, 0, 2,
    {}), (30, 0, 2, {})])
val v_filter_2: N = vector(filter(0, 2)[(3, 0, 1 {0 :: 1}), (30, 0, 2,
    {}), (30, 0, 2, {})])
val v_filter_3: N = vector(filter(0, 1)[(3, 0, 1 {0 :: 1}), (30, 0, 2,
    {}), (30, 0, 2, {})])
val mask_1: N = mask([(4, 0, 3), (32, 0, 30), (32, 0, 30)])
val const_neg1: N = const(-1)
var pt2: P[3] = 0
pt2[0] = v_filter_1
pt2[1] = v_filter_3
pt2[2] = v_filter_2
var __partial_1: N[3] = 0
for i0 in range(3) {
    instr2 = mul(N, mask_1, pt2[i0])
    __partial_1[i0] = instr2
}
encode(const_neg1)
for i0 in range(3) {
    encode(__partial_1[i0])
}
var __out: C = 0
var __reduce_1: C = 0
for i0 in range(3) {
    instr4 = rot(CC, (i0 * -1), v_img_1)
    instr6 = mul(CP, instr4, __partial_1[i0])
    instr7 = add(CC, __reduce_1, instr6)
    __reduce_1 = instr7
}
instr9 = rot(CC, -2048, __reduce_1)
instr10 = add(CC, __reduce_1, instr9)
instr11 = rot(CC, -1024, instr10)
instr12 = add(CC, instr11, instr10)
```

```
__out = instr12
```

### distance e1-o0

```
val v_point_1: C = vector(point(0)[(64, 0, 0 {0 :: 1})])
val v_tests_1: N = vector(tests.Roll(1,0)(0, 29)[(64, 0, 0 {0 :: 1})])
val v_tests_2: N = vector(tests.Roll(1,0)(0, 27)[(64, 0, 0 {0 :: 1})])
val v_tests_3: N = vector(tests.Roll(1,0)(0, 44)[(64, 0, 0 {0 :: 1})])
val v_tests_4: N = vector(tests.Roll(1,0)(0, 55)[(64, 0, 0 {0 :: 1})])
val v_tests_5: N = vector(tests.Roll(1,0)(0, 57)[(64, 0, 0 {0 :: 1})])
val v_tests_6: N = vector(tests.Roll(1,0)(0, 2)[(64, 0, 0 {0 :: 1})])
val v_tests_7: N = vector(tests.Roll(1,0)(0, 26)[(64, 0, 0 {0 :: 1})])
val v_tests_8: N = vector(tests.Roll(1,0)(0, 35)[(64, 0, 0 {0 :: 1})])
val v_tests_9: N = vector(tests.Roll(1,0)(0, 61)[(64, 0, 0 {0 :: 1})])
val v_tests_10: N = vector(tests.Roll(1,0)(0, 0)[(64, 0, 0 {0 :: 1})])
val v_tests_11: N = vector(tests.Roll(1,0)(0, 9)[(64, 0, 0 {0 :: 1})])
val v_tests_12: N = vector(tests.Roll(1,0)(0, 17)[(64, 0, 0 {0 :: 1})])
val v_tests_13: N = vector(tests.Roll(1,0)(0, 20)[(64, 0, 0 {0 :: 1})])
val v_tests_14: N = vector(tests.Roll(1,0)(0, 12)[(64, 0, 0 {0 :: 1})])
val v_tests_15: N = vector(tests.Roll(1,0)(0, 22)[(64, 0, 0 {0 :: 1})])
val v_tests_16: N = vector(tests.Roll(1,0)(0, 11)[(64, 0, 0 {0 :: 1})])
val v_tests_17: N = vector(tests.Roll(1,0)(0, 18)[(64, 0, 0 {0 :: 1})])
val v_tests_18: N = vector(tests.Roll(1,0)(0, 60)[(64, 0, 0 {0 :: 1})])
val v_tests_19: N = vector(tests.Roll(1,0)(0, 54)[(64, 0, 0 {0 :: 1})])
val v_tests_20: N = vector(tests.Roll(1,0)(0, 63)[(64, 0, 0 {0 :: 1})])
val v_tests_21: N = vector(tests.Roll(1,0)(0, 38)[(64, 0, 0 {0 :: 1})])
val v_tests_22: N = vector(tests.Roll(1,0)(0, 7)[(64, 0, 0 {0 :: 1})])
val v_tests_23: N = vector(tests.Roll(1,0)(0, 16)[(64, 0, 0 {0 :: 1})])
val v_tests_24: N = vector(tests.Roll(1,0)(0, 5)[(64, 0, 0 {0 :: 1})])
val v_tests_25: N = vector(tests.Roll(1,0)(0, 21)[(64, 0, 0 {0 :: 1})])
val v_tests_26: N = vector(tests.Roll(1,0)(0, 47)[(64, 0, 0 {0 :: 1})])
val v_tests_27: N = vector(tests.Roll(1,0)(0, 24)[(64, 0, 0 {0 :: 1})])
val v_tests_28: N = vector(tests.Roll(1,0)(0, 43)[(64, 0, 0 {0 :: 1})])
val v_tests_29: N = vector(tests.Roll(1,0)(0, 48)[(64, 0, 0 {0 :: 1})])
val v_tests_30: N = vector(tests.Roll(1,0)(0, 36)[(64, 0, 0 {0 :: 1})])
val v_tests_31: N = vector(tests.Roll(1,0)(0, 53)[(64, 0, 0 {0 :: 1})])
val v_tests_32: N = vector(tests.Roll(1,0)(0, 32)[(64, 0, 0 {0 :: 1})])
val v_tests_33: N = vector(tests.Roll(1,0)(0, 1)[(64, 0, 0 {0 :: 1})])
val v_tests_34: N = vector(tests.Roll(1,0)(0, 3)[(64, 0, 0 {0 :: 1})])
val v_tests_35: N = vector(tests.Roll(1,0)(0, 30)[(64, 0, 0 {0 :: 1})])
val v_tests_36: N = vector(tests.Roll(1,0)(0, 42)[(64, 0, 0 {0 :: 1})])
val v_tests_37: N = vector(tests.Roll(1,0)(0, 59)[(64, 0, 0 {0 :: 1})])
val v_tests_38: N = vector(tests.Roll(1,0)(0, 6)[(64, 0, 0 {0 :: 1})])
val v_tests_39: N = vector(tests.Roll(1,0)(0, 13)[(64, 0, 0 {0 :: 1})])
val v_tests_40: N = vector(tests.Roll(1,0)(0, 15)[(64, 0, 0 {0 :: 1})])
val v_tests_41: N = vector(tests.Roll(1,0)(0, 40)[(64, 0, 0 {0 :: 1})])
val v_tests_42: N = vector(tests.Roll(1,0)(0, 51)[(64, 0, 0 {0 :: 1})])
val v_tests_43: N = vector(tests.Roll(1,0)(0, 8)[(64, 0, 0 {0 :: 1})])
val v_tests_44: N = vector(tests.Roll(1,0)(0, 37)[(64, 0, 0 {0 :: 1})])
```

```
val v_tests_45: N = vector(tests.Roll(1,0)(0, 46)[(64, 0, 0 {0 :: 1})])
val v_tests_46: N = vector(tests.Roll(1,0)(0, 56)[(64, 0, 0 {0 :: 1})])
val v_tests_47: N = vector(tests.Roll(1,0)(0, 39)[(64, 0, 0 {0 :: 1})])
val v_tests_48: N = vector(tests.Roll(1,0)(0, 58)[(64, 0, 0 {0 :: 1})])
val v_tests_49: N = vector(tests.Roll(1,0)(0, 25)[(64, 0, 0 {0 :: 1})])
val v_tests_50: N = vector(tests.Roll(1,0)(0, 62)[(64, 0, 0 {0 :: 1})])
val v_tests_51: N = vector(tests.Roll(1,0)(0, 28)[(64, 0, 0 {0 :: 1})])
val v_tests_52: N = vector(tests.Roll(1,0)(0, 34)[(64, 0, 0 {0 :: 1})])
val v_tests_53: N = vector(tests.Roll(1,0)(0, 23)[(64, 0, 0 {0 :: 1})])
val v_tests_54: N = vector(tests.Roll(1,0)(0, 10)[(64, 0, 0 {0 :: 1})])
val v_tests_55: N = vector(tests.Roll(1,0)(0, 14)[(64, 0, 0 {0 :: 1})])
val v_tests_56: N = vector(tests.Roll(1,0)(0, 33)[(64, 0, 0 {0 :: 1})])
val v_tests_57: N = vector(tests.Roll(1,0)(0, 4)[(64, 0, 0 {0 :: 1})])
val v_tests_58: N = vector(tests.Roll(1,0)(0, 41)[(64, 0, 0 {0 :: 1})])
val v_tests_59: N = vector(tests.Roll(1,0)(0, 50)[(64, 0, 0 {0 :: 1})])
val v_tests_60: N = vector(tests.Roll(1,0)(0, 52)[(64, 0, 0 {0 :: 1})])
val v_tests_61: N = vector(tests.Roll(1,0)(0, 49)[(64, 0, 0 {0 :: 1})])
val v_tests_62: N = vector(tests.Roll(1,0)(0, 19)[(64, 0, 0 {0 :: 1})])
val v_tests_63: N = vector(tests.Roll(1,0)(0, 45)[(64, 0, 0 {0 :: 1})])
val v_tests_64: N = vector(tests.Roll(1,0)(0, 31)[(64, 0, 0 {0 :: 1})])
val const_neg1: N = const(-1)
encode(v_tests_19)
encode(v_tests_48)
encode(v_tests_52)
encode(v_tests_41)
encode(v_tests_54)
encode(v_tests_42)
encode(v_tests_13)
encode(v_tests_4)
encode(v_tests_11)
encode(v_tests_38)
encode(v_tests_15)
encode(v_tests_53)
encode(v_tests_32)
encode(v_tests_29)
encode(v_tests_18)
encode(v_tests_27)
encode(v_tests_58)
encode(v_tests_47)
encode(v_tests_8)
encode(v_tests_63)
encode(v_tests_7)
encode(v_tests_62)
encode(v_tests_16)
encode(v_tests_31)
encode(v_tests_45)
encode(v_tests_50)
encode(v_tests_20)
```

```
encode(v_tests_56)
encode(v_tests_2)
encode(v_tests_43)
encode(v_tests_10)
encode(v_tests_57)
encode(v_tests_6)
encode(v_tests_34)
encode(v_tests_17)
encode(v_tests_61)
encode(v_tests_3)
encode(v_tests_39)
encode(v_tests_26)
encode(v_tests_35)
encode(v_tests_25)
encode(v_tests_46)
encode(v_tests_36)
encode(v_tests_49)
encode(v_tests_22)
encode(v_tests_23)
encode(v_tests_28)
encode(v_tests_30)
encode(v_tests_12)
encode(v_tests_59)
encode(v_tests_40)
encode(v_tests_33)
encode(v_tests_24)
encode(v_tests_1)
encode(v_tests_44)
encode(v_tests_60)
encode(v_tests_21)
encode(v_tests_37)
encode(v_tests_64)
encode(v_tests_51)
encode(v_tests_9)
encode(v_tests_14)
encode(v_tests_55)
encode(v_tests_5)
encode(const_neg1)
var pt1: P[64] = 0
pt1[0] = v_tests_10
pt1[1] = v_tests_33
pt1[2] = v_tests_6
pt1[3] = v_tests_34
pt1[4] = v_tests_57
pt1[5] = v_tests_24
pt1[6] = v_tests_38
pt1[7] = v_tests_22
pt1[8] = v_tests_43
```

```
pt1[9]  = v_tests_11
pt1[10] = v_tests_54
pt1[11] = v_tests_16
pt1[12] = v_tests_14
pt1[13] = v_tests_39
pt1[14] = v_tests_55
pt1[15] = v_tests_40
pt1[16] = v_tests_23
pt1[17] = v_tests_12
pt1[18] = v_tests_17
pt1[19] = v_tests_62
pt1[20] = v_tests_13
pt1[21] = v_tests_25
pt1[22] = v_tests_15
pt1[23] = v_tests_53
pt1[24] = v_tests_27
pt1[25] = v_tests_49
pt1[26] = v_tests_7
pt1[27] = v_tests_2
pt1[28] = v_tests_51
pt1[29] = v_tests_1
pt1[30] = v_tests_35
pt1[31] = v_tests_64
pt1[32] = v_tests_32
pt1[33] = v_tests_56
pt1[34] = v_tests_52
pt1[35] = v_tests_8
pt1[36] = v_tests_30
pt1[37] = v_tests_44
pt1[38] = v_tests_21
pt1[39] = v_tests_47
pt1[40] = v_tests_41
pt1[41] = v_tests_58
pt1[42] = v_tests_36
pt1[43] = v_tests_28
pt1[44] = v_tests_3
pt1[45] = v_tests_63
pt1[46] = v_tests_45
pt1[47] = v_tests_26
pt1[48] = v_tests_29
pt1[49] = v_tests_61
pt1[50] = v_tests_59
pt1[51] = v_tests_42
pt1[52] = v_tests_60
pt1[53] = v_tests_31
pt1[54] = v_tests_19
pt1[55] = v_tests_4
pt1[56] = v_tests_46
```

```
pt1[57] = v_tests_5
pt1[58] = v_tests_48
pt1[59] = v_tests_37
pt1[60] = v_tests_18
pt1[61] = v_tests_9
pt1[62] = v_tests_50
pt1[63] = v_tests_20
var __out: C = 0
var __reduce_1: C[64] = 0
for i6 in range(64) {
    instr1 = rot(CC, (0 + i6), v_point_1)
    instr3 = sub(CP, instr1, pt1[i6])
    instr4 = mul(CC, instr3, instr3)
    __reduce_1[i6] = instr4
}
instr6 = add(CC, __reduce_1[21], __reduce_1[20])
instr7 = add(CC, __reduce_1[23], __reduce_1[22])
instr8 = add(CC, instr6, instr7)
instr9 = add(CC, __reduce_1[17], __reduce_1[16])
instr10 = add(CC, __reduce_1[19], __reduce_1[18])
instr11 = add(CC, instr9, instr10)
instr12 = add(CC, instr8, instr11)
instr13 = add(CC, __reduce_1[29], __reduce_1[28])
instr14 = add(CC, __reduce_1[31], __reduce_1[30])
instr15 = add(CC, instr13, instr14)
instr16 = add(CC, __reduce_1[25], __reduce_1[24])
instr17 = add(CC, __reduce_1[27], __reduce_1[26])
instr18 = add(CC, instr16, instr17)
instr19 = add(CC, instr15, instr18)
instr20 = add(CC, instr12, instr19)
instr21 = add(CC, __reduce_1[5], __reduce_1[4])
instr22 = add(CC, __reduce_1[7], __reduce_1[6])
instr23 = add(CC, instr21, instr22)
instr24 = add(CC, __reduce_1[1], __reduce_1[0])
instr25 = add(CC, __reduce_1[3], __reduce_1[2])
instr26 = add(CC, instr24, instr25)
instr27 = add(CC, instr23, instr26)
instr28 = add(CC, __reduce_1[13], __reduce_1[12])
instr29 = add(CC, __reduce_1[15], __reduce_1[14])
instr30 = add(CC, instr28, instr29)
instr31 = add(CC, __reduce_1[9], __reduce_1[8])
instr32 = add(CC, __reduce_1[11], __reduce_1[10])
instr33 = add(CC, instr31, instr32)
instr34 = add(CC, instr30, instr33)
instr35 = add(CC, instr27, instr34)
instr36 = add(CC, instr20, instr35)
instr37 = add(CC, __reduce_1[53], __reduce_1[52])
instr38 = add(CC, __reduce_1[55], __reduce_1[54])
```

```
instr39 = add(CC, instr37, instr38)
instr40 = add(CC, __reduce_1[49], __reduce_1[48])
instr41 = add(CC, __reduce_1[51], __reduce_1[50])
instr42 = add(CC, instr40, instr41)
instr43 = add(CC, instr39, instr42)
instr44 = add(CC, __reduce_1[61], __reduce_1[60])
instr45 = add(CC, __reduce_1[63], __reduce_1[62])
instr46 = add(CC, instr44, instr45)
instr47 = add(CC, __reduce_1[57], __reduce_1[56])
instr48 = add(CC, __reduce_1[59], __reduce_1[58])
instr49 = add(CC, instr47, instr48)
instr50 = add(CC, instr46, instr49)
instr51 = add(CC, instr43, instr50)
instr52 = add(CC, __reduce_1[37], __reduce_1[36])
instr53 = add(CC, __reduce_1[39], __reduce_1[38])
instr54 = add(CC, instr52, instr53)
instr55 = add(CC, __reduce_1[33], __reduce_1[32])
instr56 = add(CC, __reduce_1[35], __reduce_1[34])
instr57 = add(CC, instr55, instr56)
instr58 = add(CC, instr54, instr57)
instr59 = add(CC, __reduce_1[45], __reduce_1[44])
instr60 = add(CC, __reduce_1[47], __reduce_1[46])
instr61 = add(CC, instr59, instr60)
instr62 = add(CC, __reduce_1[41], __reduce_1[40])
instr63 = add(CC, __reduce_1[43], __reduce_1[42])
instr64 = add(CC, instr62, instr63)
instr65 = add(CC, instr61, instr64)
instr66 = add(CC, instr58, instr65)
instr67 = add(CC, instr51, instr66)
instr68 = add(CC, instr36, instr67)
__out = instr68
```

**distance e2-o0**

```
val v_point_1: C = vector(point(0)[(64, 0, 0, {}), (32, 0, 0 {0 :: 2})])
val v_point_2: C = vector(point(1)[(64, 0, 0, {}), (32, 0, 0 {0 :: 2})])
val v_tests_1: N = vector(tests(0, 0)[(64, 0, 0 {0 :: 1}), (32, 0, 0 {1
    :: 2})])
val v_tests_2: N = vector(tests(0, 1)[(64, 0, 0 {0 :: 1}), (32, 0, 0 {1
    :: 2})])
val const_neg1: N = const(-1)
encode(v_tests_2)
encode(v_tests_1)
encode(const_neg1)
var ct1: C[2] = 0
ct1[0] = v_point_1
ct1[1] = v_point_2
var pt1: P[2] = 0
pt1[0] = v_tests_1
```

```
pt1[1] = v_tests_2
var __out: C = 0
var __reduce_1: C = 0
for i3i in range(2) {
    instr2 = sub(CP, ct1[i3i], pt1[i3i])
    instr3 = mul(CC, instr2, instr2)
    instr4 = add(CC, __reduce_1, instr3)
    __reduce_1 = instr4
}
instr6 = rot(CC, -16, __reduce_1)
instr7 = add(CC, __reduce_1, instr6)
instr8 = rot(CC, -8, instr7)
instr9 = add(CC, instr7, instr8)
instr10 = rot(CC, -4, instr9)
instr11 = add(CC, instr9, instr10)
instr12 = rot(CC, -2, instr11)
instr13 = add(CC, instr11, instr12)
instr14 = rot(CC, -1, instr13)
instr15 = add(CC, instr13, instr14)
__out = instr15
```

**double-matmul e1-o0**

```
val v_B_1: C = vector(B(0, 0)[(16, 0, 0 {1 :: 1}), (16, 0, 0 {0 :: 1}),
    (16, 0, 0, {})])
val v_A2_1: N = vector(A2(0, 0)[(16, 0, 0, {}), (16, 0, 0 {0 :: 1}), (16,
    0, 0 {1 :: 1})])
val v_A1_1: N = vector(A1(0, 0)[(16, 0, 0, {}), (16, 0, 0 {1 :: 1}), (16,
    0, 0 {0 :: 1})])
val mask_1: N = mask([(16, 0, 15), (16, 0, 0), (16, 0, 15)])
val const_neg1: N = const(-1)
encode(v_A2_1)
encode(v_A1_1)
encode(mask_1)
encode(const_neg1)
var res: C = 0
instr2 = mul(CP, v_B_1, v_A1_1)
instr3 = rot(CC, -128, instr2)
instr4 = add(CC, instr2, instr3)
instr5 = rot(CC, -64, instr4)
instr6 = add(CC, instr4, instr5)
instr7 = rot(CC, -32, instr6)
instr8 = add(CC, instr6, instr7)
instr9 = rot(CC, -16, instr8)
instr10 = add(CC, instr8, instr9)
res = instr10
var ct2: C[1] = 0
ct2[0] = res
var __circ_1: C[1] = 0
```

```
for i in range(1) {
    instr13 = mul(CP, ct2[i], mask_1)
    instr14 = rot(CC, 16, instr13)
    instr15 = add(CC, instr13, instr14)
    instr16 = rot(CC, 32, instr15)
    instr17 = add(CC, instr15, instr16)
    instr18 = rot(CC, 64, instr17)
    instr19 = add(CC, instr17, instr18)
    instr20 = rot(CC, 128, instr19)
    instr21 = add(CC, instr19, instr20)
    __circ_1[i] = instr21
}
var __out: C = 0
instr24 = mul(CP, __circ_1[0], v_A2_1)
instr25 = rot(CC, -8, instr24)
instr26 = add(CC, instr24, instr25)
instr27 = rot(CC, -4, instr26)
instr28 = add(CC, instr26, instr27)
instr29 = rot(CC, -2, instr28)
instr30 = add(CC, instr28, instr29)
instr31 = rot(CC, -1, instr30)
instr32 = add(CC, instr30, instr31)
__out = instr32
```

**retrieval-256 e1-o0**

```
val v_values_1: C = vector(values(0)[(256, 0, 0 {0 :: 1})])
val v_keys_1: C = vector(keys(0, 0)[(8, 0, 0 {1 :: 1}), (256, 0, 0 {0 ::
    1})])
val v_query_1: C = vector(query(0)[(8, 0, 0 {0 :: 1}), (256, 0, 0, {})])
val const_1: N = const(1)
val const_neg1: N = const(-1)
encode(const_1)
encode(const_neg1)
var mask: C = 0
instr3 = sub(CC, v_query_1, v_keys_1)
instr4 = mul(CC, instr3, instr3)
instr5 = mul(CP, instr4, const_neg1)
instr6 = add(CP, instr5, const_1)
instr7 = rot(CC, -1024, instr6)
instr8 = mul(CC, instr6, instr7)
instr9 = rot(CC, -512, instr8)
instr10 = mul(CC, instr8, instr9)
instr11 = rot(CC, -256, instr10)
instr12 = mul(CC, instr10, instr11)
mask = instr12
var __out: C = 0
instr15 = mul(CC, v_values_1, mask)
instr16 = rot(CC, -128, instr15)
```

```
instr17 = add(CC, instr15, instr16)
instr18 = rot(CC, -64, instr17)
instr19 = add(CC, instr17, instr18)
instr20 = rot(CC, -32, instr19)
instr21 = add(CC, instr19, instr20)
instr22 = rot(CC, -16, instr21)
instr23 = add(CC, instr21, instr22)
instr24 = rot(CC, -8, instr23)
instr25 = add(CC, instr23, instr24)
instr26 = rot(CC, -4, instr25)
instr27 = add(CC, instr25, instr26)
instr28 = rot(CC, -2, instr27)
instr29 = add(CC, instr27, instr28)
instr30 = rot(CC, -1, instr29)
instr31 = add(CC, instr29, instr30)
__out = instr31
```

**retrieval-1024 e1-o0**

```
val v_query_1: C = vector(query(6)[(1024, 0, 0, {})])
val v_query_2: C = vector(query(9)[(1024, 0, 0, {})])
val v_query_3: C = vector(query(2)[(1024, 0, 0, {})])
val v_query_4: C = vector(query(5)[(1024, 0, 0, {})])
val v_query_5: C = vector(query(7)[(1024, 0, 0, {})])
val v_keys_1: C = vector(keys(0, 8)[(1024, 0, 0 {0 :: 1})])
val v_query_6: C = vector(query(4)[(1024, 0, 0, {})])
val v_keys_2: C = vector(keys(0, 6)[(1024, 0, 0 {0 :: 1})])
val v_keys_3: C = vector(keys(0, 9)[(1024, 0, 0 {0 :: 1})])
val v_query_7: C = vector(query(8)[(1024, 0, 0, {})])
val v_query_8: C = vector(query(3)[(1024, 0, 0, {})])
val v_keys_4: C = vector(keys(0, 1)[(1024, 0, 0 {0 :: 1})])
val v_keys_5: C = vector(keys(0, 7)[(1024, 0, 0 {0 :: 1})])
val v_query_9: C = vector(query(1)[(1024, 0, 0, {})])
val v_query_10: C = vector(query(0)[(1024, 0, 0, {})])
val v_keys_6: C = vector(keys(0, 5)[(1024, 0, 0 {0 :: 1})])
val v_keys_7: C = vector(keys(0, 0)[(1024, 0, 0 {0 :: 1})])
val v_keys_8: C = vector(keys(0, 2)[(1024, 0, 0 {0 :: 1})])
val v_keys_9: C = vector(keys(0, 3)[(1024, 0, 0 {0 :: 1})])
val v_keys_10: C = vector(keys(0, 4)[(1024, 0, 0 {0 :: 1})])
val v_values_1: C = vector(values(0)[(1024, 0, 0 {0 :: 1})])
val const_1: N = const(1)
val const_neg1: N = const(-1)
encode(const_1)
encode(const_neg1)
var ct1: C[10] = 0
ct1[0] = v_query_10
ct1[1] = v_query_9
ct1[2] = v_query_3
ct1[3] = v_query_8
```

```
ct1[4] = v_query_6
ct1[5] = v_query_4
ct1[6] = v_query_1
ct1[7] = v_query_5
ct1[8] = v_query_7
ct1[9] = v_query_2
var ct2: C[10] = 0
ct2[0] = v_keys_7
ct2[1] = v_keys_4
ct2[2] = v_keys_8
ct2[3] = v_keys_9
ct2[4] = v_keys_10
ct2[5] = v_keys_6
ct2[6] = v_keys_2
ct2[7] = v_keys_5
ct2[8] = v_keys_1
ct2[9] = v_keys_3
var mask: C = 0
var __reduce_1: C[10] = 1
for i5 in range(10) {
    instr3 = sub(CC, ct1[i5], ct2[i5])
    instr4 = mul(CC, instr3, instr3)
    instr5 = mul(CP, instr4, const_neg1)
    instr6 = add(CP, instr5, const_1)
    __reduce_1[i5] = instr6
}
instr8 = mul(CC, __reduce_1[1], __reduce_1[0])
instr9 = mul(CC, __reduce_1[7], __reduce_1[6])
instr10 = mul(CC, __reduce_1[9], __reduce_1[8])
instr11 = mul(CC, instr9, instr10)
instr12 = mul(CC, __reduce_1[3], __reduce_1[2])
instr13 = mul(CC, __reduce_1[5], __reduce_1[4])
instr14 = mul(CC, instr12, instr13)
instr15 = mul(CC, instr11, instr14)
instr16 = mul(CC, instr8, instr15)
mask = instr16
var __out: C = 0
instr19 = mul(CC, v_values_1, mask)
instr20 = rot(CC, -512, instr19)
instr21 = add(CC, instr19, instr20)
instr22 = rot(CC, -256, instr21)
instr23 = add(CC, instr21, instr22)
instr24 = rot(CC, -128, instr23)
instr25 = add(CC, instr23, instr24)
instr26 = rot(CC, -64, instr25)
instr27 = add(CC, instr25, instr26)
instr28 = rot(CC, -32, instr27)
instr29 = add(CC, instr27, instr28)
```

```
instr30 = rot(CC, -16, instr29)
instr31 = add(CC, instr29, instr30)
instr32 = rot(CC, -8, instr31)
instr33 = add(CC, instr31, instr32)
instr34 = rot(CC, -4, instr33)
instr35 = add(CC, instr33, instr34)
instr36 = rot(CC, -2, instr35)
instr37 = add(CC, instr35, instr36)
instr38 = rot(CC, -1, instr37)
instr39 = add(CC, instr37, instr38)
__out = instr39
```

**retrieval-1024 e2-o0**

```
val v_values_1: C = vector(values(0)[(1024, 0, 0 {0 :: 1})])
val v_query_1: C = vector(query(0)[(8, 0, 0 {0 :: 1}), (1024, 0, 0, {})])
val v_keys_1: C = vector(keys(0, 8)[(2, 0, 6 {1 :: 1}), (1024, 0, 0 {0 ::
    1})])
val v_query_2: C = vector(query(8)[(2, 0, 6 {0 :: 1}), (1024, 0, 0, {})])
val v_keys_2: C = vector(keys(0, 0)[(8, 0, 0 {1 :: 1}), (1024, 0, 0 {0 ::
    1})])
val const_1: N = const(1)
val const_neg1: N = const(-1)
encode(const_1)
encode(const_neg1)
var ct2: C[2] = 0
ct2[0] = v_keys_2
ct2[1] = v_keys_1
var ct1: C[2] = 0
ct1[0] = v_query_1
ct1[1] = v_query_2
var mask: C = 0
var __reduce_1: C = 1
for i4o in range(2) {
    instr3 = sub(CC, ct1[i4o], ct2[i4o])
    instr4 = mul(CC, instr3, instr3)
    instr5 = mul(CP, instr4, const_neg1)
    instr6 = add(CP, instr5, const_1)
    instr7 = mul(CC, __reduce_1, instr6)
    __reduce_1 = instr7
}
instr9 = rot(CC, -4096, __reduce_1)
instr10 = mul(CC, __reduce_1, instr9)
instr11 = rot(CC, -2048, instr10)
instr12 = mul(CC, instr10, instr11)
instr13 = rot(CC, -1024, instr12)
instr14 = mul(CC, instr12, instr13)
mask = instr14
var __out: C = 0
```

```
instr17 = mul(CC, v_values_1, mask)
instr18 = rot(CC, -512, instr17)
instr19 = add(CC, instr17, instr18)
instr20 = rot(CC, -256, instr19)
instr21 = add(CC, instr19, instr20)
instr22 = rot(CC, -128, instr21)
instr23 = add(CC, instr21, instr22)
instr24 = rot(CC, -64, instr23)
instr25 = add(CC, instr23, instr24)
instr26 = rot(CC, -32, instr25)
instr27 = add(CC, instr25, instr26)
instr28 = rot(CC, -16, instr27)
instr29 = add(CC, instr27, instr28)
instr30 = rot(CC, -8, instr29)
instr31 = add(CC, instr29, instr30)
instr32 = rot(CC, -4, instr31)
instr33 = add(CC, instr31, instr32)
instr34 = rot(CC, -2, instr33)
instr35 = add(CC, instr33, instr34)
instr36 = rot(CC, -1, instr35)
instr37 = add(CC, instr35, instr36)
__out = instr37
```

**set-union-16 e1-o0**

```
val v_b_data_1: C = vector(b_data(0)[(16, 0, 0 {0 :: 1})])
val v_a_id_1: C = vector(a_id(0, 0)[(16, 0, 0 {0 :: 1}), (4, 0, 0 {1 ::
    1}), (16, 0, 0, {})])
val v_b_id_1: C = vector(b_id(0, 0)[(16, 0, 0, {}), (4, 0, 0 {1 :: 1}),
    (16, 0, 0 {0 :: 1})])
val v_a_data_1: C = vector(a_data(0)[(16, 0, 0 {0 :: 1})])
val const_1: N = const(1)
val const_neg1: N = const(-1)
encode(const_1)
encode(const_neg1)
var b_sum: C = 0
instr4 = sub(CC, v_a_id_1, v_b_id_1)
instr5 = mul(CC, instr4, instr4)
instr6 = mul(CP, instr5, const_neg1)
instr7 = add(CP, instr6, const_1)
instr8 = rot(CC, -32, instr7)
instr9 = mul(CC, instr7, instr8)
instr10 = rot(CC, -16, instr9)
instr11 = mul(CC, instr9, instr10)
instr12 = mul(CP, instr11, const_neg1)
instr13 = add(CP, instr12, const_1)
instr14 = rot(CC, -512, instr13)
instr15 = mul(CC, instr13, instr14)
instr16 = rot(CC, -256, instr15)
```

```
instr17 = mul(CC, instr15, instr16)
instr18 = rot(CC, -128, instr17)
instr19 = mul(CC, instr17, instr18)
instr20 = rot(CC, -64, instr19)
instr21 = mul(CC, instr19, instr20)
instr22 = mul(CC, v_b_data_1, instr21)
instr23 = rot(CC, -8, instr22)
instr24 = add(CC, instr22, instr23)
instr25 = rot(CC, -4, instr24)
instr26 = add(CC, instr24, instr25)
instr27 = rot(CC, -2, instr26)
instr28 = add(CC, instr26, instr27)
instr29 = rot(CC, -1, instr28)
instr30 = add(CC, instr28, instr29)
b_sum = instr30
var a_sum: C = 0
instr32 = rot(CC, -8, v_a_data_1)
instr33 = add(CC, v_a_data_1, instr32)
instr34 = rot(CC, -4, instr33)
instr35 = add(CC, instr33, instr34)
instr36 = rot(CC, -2, instr35)
instr37 = add(CC, instr35, instr36)
instr38 = rot(CC, -1, instr37)
instr39 = add(CC, instr37, instr38)
a_sum = instr39
var __out: C = 0
instr42 = add(CC, a_sum, b_sum)
__out = instr42
```

**set-union-128**

```
val v_a_id_1: C = vector(a_id(0, 2)[(128, 0, 0 {0 :: 1}), (128, 0, 0, {})
   ])
val v_b_id_1: C = vector(b_id(0, 1)[(128, 0, 0, {}), (128, 0, 0 {0 :: 1})
   ])
val v_b_id_2: C = vector(b_id(0, 6)[(128, 0, 0, {}), (128, 0, 0 {0 :: 1})
   ])
val v_b_id_3: C = vector(b_id(0, 2)[(128, 0, 0, {}), (128, 0, 0 {0 :: 1})
   ])
val v_a_id_2: C = vector(a_id(0, 0)[(128, 0, 0 {0 :: 1}), (128, 0, 0, {})
   ])
val v_a_data_1: C = vector(a_data(0)[(128, 0, 0 {0 :: 1})])
val v_b_data_1: C = vector(b_data(0)[(128, 0, 0 {0 :: 1})])
val v_a_id_3: C = vector(a_id(0, 4)[(128, 0, 0 {0 :: 1}), (128, 0, 0, {})
   ])
val v_a_id_4: C = vector(a_id(0, 6)[(128, 0, 0 {0 :: 1}), (128, 0, 0, {})
   ])
val v_b_id_4: C = vector(b_id(0, 0)[(128, 0, 0, {}), (128, 0, 0 {0 :: 1})
   ])
```

```
val v_b_id_5: C = vector(b_id(0, 3)[(128, 0, 0, {}), (128, 0, 0 {0 :: 1})
    ])
val v_b_id_6: C = vector(b_id(0, 4)[(128, 0, 0, {}), (128, 0, 0 {0 :: 1})
    ])
val v_a_id_5: C = vector(a_id(0, 1)[(128, 0, 0 {0 :: 1}), (128, 0, 0, {})
    ])
val v_b_id_7: C = vector(b_id(0, 5)[(128, 0, 0, {}), (128, 0, 0 {0 :: 1})
    ])
val v_a_id_6: C = vector(a_id(0, 3)[(128, 0, 0 {0 :: 1}), (128, 0, 0, {})
    ])
val v_a_id_7: C = vector(a_id(0, 5)[(128, 0, 0 {0 :: 1}), (128, 0, 0, {})
    ])
val const_1: N = const(1)
val const_neg1: N = const(-1)
encode(const_1)
encode(const_neg1)
var ct2: C[7] = 0
ct2[0] = v_a_id_2
ct2[1] = v_a_id_5
ct2[2] = v_a_id_1
ct2[3] = v_a_id_6
ct2[4] = v_a_id_3
ct2[5] = v_a_id_7
ct2[6] = v_a_id_4
var ct3: C[7] = 0
ct3[0] = v_b_id_4
ct3[1] = v_b_id_1
ct3[2] = v_b_id_3
ct3[3] = v_b_id_5
ct3[4] = v_b_id_6
ct3[5] = v_b_id_7
ct3[6] = v_b_id_2
var b_sum: C = 0
var __reduce_1: C[7] = 1
for i8 in range(7) {
    instr4 = sub(CC, ct2[i8], ct3[i8])
    instr5 = mul(CC, instr4, instr4)
    instr6 = mul(CP, instr5, const_neg1)
    instr7 = add(CP, instr6, const_1)
    __reduce_1[i8] = instr7
}
instr9 = mul(CC, __reduce_1[1], __reduce_1[0])
instr10 = mul(CC, __reduce_1[3], __reduce_1[2])
instr11 = mul(CC, __reduce_1[5], __reduce_1[4])
instr12 = mul(CC, instr10, instr11)
instr13 = mul(CC, instr9, instr12)
instr14 = mul(CC, __reduce_1[6], instr13)
instr15 = mul(CP, instr14, const_neg1)
```

```
instr16 = add(CP, instr15, const_1)
instr17 = rot(CC, -8192, instr16)
instr18 = mul(CC, instr16, instr17)
instr19 = rot(CC, -4096, instr18)
instr20 = mul(CC, instr18, instr19)
instr21 = rot(CC, -2048, instr20)
instr22 = mul(CC, instr20, instr21)
instr23 = rot(CC, -1024, instr22)
instr24 = mul(CC, instr22, instr23)
instr25 = rot(CC, -512, instr24)
instr26 = mul(CC, instr24, instr25)
instr27 = rot(CC, -256, instr26)
instr28 = mul(CC, instr26, instr27)
instr29 = rot(CC, -128, instr28)
instr30 = mul(CC, instr28, instr29)
instr31 = mul(CC, v_b_data_1, instr30)
instr32 = rot(CC, -64, instr31)
instr33 = add(CC, instr31, instr32)
instr34 = rot(CC, -32, instr33)
instr35 = add(CC, instr33, instr34)
instr36 = rot(CC, -16, instr35)
instr37 = add(CC, instr35, instr36)
instr38 = rot(CC, -8, instr37)
instr39 = add(CC, instr37, instr38)
instr40 = rot(CC, -4, instr39)
instr41 = add(CC, instr39, instr40)
instr42 = rot(CC, -2, instr41)
instr43 = add(CC, instr41, instr42)
instr44 = rot(CC, -1, instr43)
instr45 = add(CC, instr43, instr44)
b_sum = instr45
var a_sum: C = 0
instr47 = rot(CC, -64, v_a_data_1)
instr48 = add(CC, v_a_data_1, instr47)
instr49 = rot(CC, -32, instr48)
instr50 = add(CC, instr48, instr49)
instr51 = rot(CC, -16, instr50)
instr52 = add(CC, instr50, instr51)
instr53 = rot(CC, -8, instr52)
instr54 = add(CC, instr52, instr53)
instr55 = rot(CC, -4, instr54)
instr56 = add(CC, instr54, instr55)
instr57 = rot(CC, -2, instr56)
instr58 = add(CC, instr56, instr57)
instr59 = rot(CC, -1, instr58)
instr60 = add(CC, instr58, instr59)
a_sum = instr60
var __out: C = 0
```

```
instr63 = add(CC, a_sum, b_sum)
__out = instr63
```